# A fully discrete energy stable scheme for a phase-field moving contact line model with variable densities and viscosities

**Guangpu Zhu[a], Huangxin Chen[b], Aifen Li[a], Shuyu Sun[c], Jun Yao [a]**

[a]*Research Center of Multiphase Flow in Porous Media, School of Petroleum Engineering, China University of Petroleum (East China), Qingdao 266580, China*
[b]*School of Mathematical Sciences and Fujian Provincial Key Laboratory on Mathematical Modeling and High Performance Scientific Computing, Xiamen University, Xiamen 361005, China*
[c]*Computational Transport Phenomena Laboratory, Division of Physical Science and Engineering, King Abdullah University of Science and Technology, Thuwal 23955-6900, Kingdom of Saudi Arabia*

## Abstract

In this work, we propose a fully discrete energy stable scheme for the phase-field moving contact line model with variable densities and viscosities. The mathematical model consists of a Cahn–Hilliard equation, a Navier–Stokes equation and the generalized Navier boundary condition for the moving contact line. A scalar auxiliary variable is adopted to transform the governing system into an equivalent form, allowing the double well potential to be treated semi-explicitly. A stabilization term is added to balance the explicit nonlinear term originating from the surface energy at fluid-solid interface. A pressure stabilization method is used to decouple the computation of velocity and pressure. Some subtle implicit-explicit treatments are adopted to deal with convention and stress terms. We establish a rigorous proof of energy stability for the proposed time-marching scheme. Then a finite difference method on staggered grids is used to spatially discretize the constructed time-marching scheme. We further prove that the fully discrete scheme also satisfies the discrete energy dissipation law. Numerical results demonstrate accuracy and energy stability of the proposed scheme. Using our numerical scheme, we analyze the contact line dynamics through a shear flow driven droplet sliding case. Three-dimensional droplet spreading is also investigated on a chemically patterned surface. Our numerical simulation accurately predicts the expected energy evolutions and it successfully reproduces expected phenomena that an oil droplet contracts inwards on a hydrophobic zone and spreads outwards quickly on a hydrophilic zone.

## 1. Introduction

The moving contact line (MCL) problem, which occurs when the interface of two immiscible fluid components touches the solid wall, is a classical problem that appears in many industrial applications [1, 2], such as spray coating, spray cooling of surfaces and enhanced oil recovery. In petroleum industry, the wettability of rock affects the efficiency of water flooding on residual oil, thus determining the oil recovery [3]. It is well known that the classical Navier–Stokes equation [4] with the no-slip boundary condition is not applicable for the MCL problem, since non-physical singularities will occur in the vicinity of the MCL [1, 2, 5]. Numerous models have been proposed over the years, aiming to understand the hydrodynamic behavior near the MCL, including molecular dynamics (MD) models [6, 7], lattice Boltzmann model [8, 9], sharp interface models

Keywords: Phase-field modelling; Moving contact line; Navier–Stokes; Energy stability.

E-mail addresses: zhugpupc@gmail.com (G. Zhu), aifenli64@gmail.com (A. Li*), shuyu.sun@kaust.edu.sa (S. Sun*)

[10-13] and the phase-field model [14-16] considered in this work. The phase-field model has been widely used in simulations of two immiscible fluid components by coupling the fourth-order Cahn–Hilliard equation with the Navier–Stokes equation via convection and stress terms [3, 17-23]. In this approach, a phase-field variable is introduced to distinguish the two immiscible phases, and the interface is treated as a thin and continuous layer to remove the singularities [24-26]. Numerous numerical schemes [27-31], which satisfy the corresponding discrete energy dissipation law, have been proposed to improve the efficiency and stability of the phase-field model.

A series of pioneering works on the phase-field MCL models can be found in [7, 18, 20, 32-35]. In this study we only focus on the generalized Navier boundary condition (GNBC). Numerous MD studies had proven that there exists a relative slip between the fluid and wall, and Qian et al. further found that the amount of slipping is proportional to the sum of tangential viscous stress and unbalanced Young stress [6, 7, 36, 37]. Subsequently, they proposed a continuum model, consisting of the Cahn–Hilliard equation, the Navier–Stokes equation and the generalized Navier boundary condition, to resolve the MCL conundrum [1]. The numerical results based on the proposed continuum model can quantitatively reproduce the MCL slip velocity profiles obtained by MD simulations [36]. Most recently, they presented a variational derivation of the GNBC through the principle of minimum energy dissipation [1, 36, 38]. This derivation reveals that the proposed slip boundary condition is consistent with the general principle of irreversible thermodynamic process [1].

From the numerical point of view, the development of energy stable schemes to solve this continuum model is quite a challenging topic. Recently, several attempts had been made in the literature to improve the efficiency and stability of numerical schemes for the phase-field MCL models. In [39], the authors proposed a least-square finite element method with a temporal discretization by the operator-splitting to overcome difficulties associated with boundary conditions and nonlinearity. Gao and Wang [40, 41] developed conditionally energy stable schemes for a phase-field MCL model using the convex splitting method. A finite difference method on the semi-staggered grids was used to discretize the nonlinear coupled system in space. In addition, the schemes based on the pressure correction method and the pressure stabilization method were compared in their work. However, a rigorous proof of the energy stability for the fully discrete scheme was not provided. Also, the implementation of nonlinear schemes produced by convex splitting method was complicated and computational costs were high. Using the stabilization approach, Shen et al. [15] developed a linear, decoupled and energy stable scheme for a matched density system with static contact line conditions, and a linear coupling scheme for dynamic contact line conditions. Aland and Chen [14] also constructed a linearly coupled scheme for a matched density system with the nonlinear potential treated by the stabilization method, and they proved the unconditional energy stability for the fully discrete finite element scheme. Similarly, using the stabilization approach, Yu and Yang [16] proposed a nonlinear coupled scheme for a non-matched density system with dynamic contact line conditions, and the spatial discretization was completed by a spectral Galerkin method. The stabilization approach mentioned above requires the second derivative of the double well potential to be bounded. Nevertheless, this is not satisfied by the double well potential. In [42], an energy stable scheme for the phase-field MCL model was constructed based on the invariant energy quadratization (IEQ) approach. The IEQ approach introduces an auxiliary variable to transform the nonlinear potential into a new form, which provides the fundamental support for the linearization treatment. However, this approach requires that the nonlinear potential is bounded from below, and this may not hold for some physical interesting models. Most recently, a scalar auxiliary variable (SAV) approach is proposed in [43], and it enjoys all advantages of the IEQ approach but overcomes most of its shortcomings [44, 45]. The SAV approach only requires that the free energy term associated with the nonlinear potential is bounded from below, which is necessary for the free energy to be physically sound [43]. In [46], we used the SAV approach to deal with the double well potential and constructed several decoupled and energy stable schemes for the hydrodynamics coupled phase-field model. Numerical results confirmed that our scheme is efficient, accurate and energy stable. However, a nonlinear surface energy potential introduced in the phase-field MCL model presents a new challenge to the development of energy stable schemes.

In this work, we adopt the combined SAV and stabilization approach to construct an energy stable time-marching scheme for the phase-field MCL model. The finite difference method on the staggered grids, i.e., also known as the MAC method [47-49], has been widely used in many engineering applications. It has been one of the simplest and effective numerical methods to solve the Navier–Stokes equation. Thus, the finite difference method on the staggered grids is implemented to spatially discretize the constructed time-marching scheme. Then we provide a rigorous proof of energy stability for the fully discrete scheme. Several two-dimensional (2D) and three-dimensional (3D) numerical experiments are conducted to validate the accuracy and energy stability of the proposed scheme.

The rest of paper is organized as follows. The coupled Cahn–Hilliard and Navier–Stokes system with the GNBC is introduced in Section 2. In Section 3 and Section 4, a time-marching scheme and a fully discrete scheme are presented and the corresponding energy estimates are given. Several numerical experiments are conducted in Section 5 and the paper is finally concluded in Section 6.

## 2. phase-field moving contact line model

In an immiscible two-phase system with the MCL, there are two types of free energy: the mixing energy controlling the interfacial dynamics [21], and the surface energy at the fluid-solid interface determining the MCL dynamics [1].

*Mixing energy*. In the phase-field model, a phase-field variable is introduced to distinguish two pure phases, and the interface is treated as a thin and continuous layer, inside which the two phases are mixed and store a mixing energy. The Ginzburg–Landau free energy functional is usually used to represent the mixing energy

$$E_{mix} = \int \left( \frac{\lambda \varepsilon}{2} |\nabla \phi|^2 + \lambda F(\phi) \right) \mathrm{d}\Omega, \tag{2.1}$$

where $\phi$ denotes the phase-field variable. It takes two constant values (often 1 and –1) in the pure phases, and varies continuously across the interface. $\varepsilon$ is a parameter associated with the interface thickness. A typical choice for $F(\phi)$ is the double well potential [15], and it can be written as

$$F(\phi) = \frac{1}{4\varepsilon} \left( \phi^2 - 1 \right)^2 . \tag{2.2}$$

$\lambda$ denotes the rescaled interfacial tension, and the scaling depends on the particular choice of $F(\phi)$ [21]. The square gradient term in (2.1) represents weakly non-local interactions between fluids that contribute to the complete mixing of phases, and the second term, bulk energy, prefers total separation of phases and produces the classical sharp-interface picture [21]. The competition between the two terms creates a diffuse interface in equilibrium [30].

*Surface energy at the fluid-solid interface.* When the diffuse interface touches solid wall, the MCL problem occurs. The angle between the fluid-fluid interface and solid wall is called "static contact angle $\theta_s$" [50, 51]. The interfacial free energy per unit area $M(\phi)$ at the fluid-solid is the function of local composition ($\theta_s$ and $\phi$). The typical surface energy at the fluid-solid surface reads:

$$E_{wf}(\phi)=\lambda\int M(\phi)d\Gamma=\int -\frac{\sqrt{2}\lambda}{3}\cos\theta_s\sin\left(\frac{\pi}{2}\phi\right)\mathrm{d}\Gamma. \tag{2.3}$$

Note that the scaling $\sqrt{2}/3$ for $\lambda$ is a consequence of the particular choice for $F(\phi)$ [21]. The total free energy of the two-phase system is the sum of the mixing energy $E_{mix}$ and the surface energy at the fluid-solid interface $E_{wf}$, and it can be written as

$$E_f(\phi)=E_{mix}(\phi)+E_{wf}(\phi)=\int\left(\frac{\lambda\varepsilon}{2}|\nabla\phi|^2+\lambda F(\phi)\right)\mathrm{d}\Omega+\lambda\int M(\phi)\mathrm{d}\Gamma. \tag{2.4}$$

Two quantities $w$ and $L$ can be defined from the variation of the total free energy $E_f$ with respect to the phase-field variable [36]

$$\delta E_f(\phi)=\int(w\delta\phi)\mathrm{d}\Omega+\int(\lambda L\delta\phi)\mathrm{d}\Gamma, \tag{2.5}$$

where $w$ is the chemical potential in the bulk

$$w:=\frac{\delta E_{mix}}{\delta\phi}=-\lambda\varepsilon\Delta\phi+\lambda F'(\phi), \tag{2.6}$$

and $L$ is the corresponding quantity at the solid wall [36],

$$L(\phi)=\varepsilon\partial_n\phi+M'(\phi). \tag{2.7}$$

Minimizing the total free energy $E_f$ with respect to $\phi$ yields the equilibrium conditions $w = C$ (constant) in the bulk and $L = 0$ at the fluid-solid interface [36].

The chemical potential gradient $\nabla w$ will arise a diffusive current in the bulk,

$$\mathbf{J}_D=-M_\phi\nabla w,$$

where $M_\phi$ is the constant mobility parameter. The diffusive current $\mathbf{J}_D$ and the material time derivative of $\phi$ satisfy the continuity equation (mass conservation law),

$$\phi_t+\nabla\cdot(\mathbf{u}\phi)=-\nabla\cdot\mathbf{J}_D=M_\phi\Delta w, \tag{2.8}$$

where $\mathbf{u}$ is the velocity. The equation (2.8) is the typical Cahn–Hilliard–type equation. The momentum equation for the hydrodynamics takes the usual form of Navier–Stokes equation

$$\rho(\mathbf{u}_t+(\mathbf{u}\cdot\nabla)\mathbf{u})+\mathbf{J}\cdot\nabla\mathbf{u}-\nabla\cdot\eta D(\mathbf{u})+\nabla p+\phi\nabla w=0, \tag{2.9}$$

$$\nabla\cdot\mathbf{u}=0, \tag{2.10}$$

where $p$ is the pressure and

$$D(\mathbf{u})=\nabla\mathbf{u}+\nabla\mathbf{u}^T,\ \mathbf{J}=\frac{\rho_1-\rho_2}{2}\mathbf{J}_D. \tag{2.11}$$

The density $\rho$ and viscosity $\eta$ satisfy the following linear relations

$$\rho=\frac{\rho_1-\rho_2}{2}\phi+\frac{\rho_1+\rho_2}{2},\ \eta=\frac{\eta_1-\eta_2}{2}\phi+\frac{\eta_1+\eta_2}{2}. \tag{2.12}$$

The following conservation property can be easily derived from (2.8), (2.10) and (2.12):

$$\rho_t + \nabla\cdot(\rho\mathbf{u}) + \nabla\cdot\mathbf{J} = 0. \tag{2.13}$$

If we consider the extra body force in the momentum equation, e.g., the gravitational force, the equation (2.9) can be modified into

$$\rho(\mathbf{u}_t + \mathbf{u}\cdot\nabla\mathbf{u}) + \mathbf{J}\cdot\nabla\mathbf{u} - \nabla\cdot\eta D(\mathbf{u}) + \nabla p + \phi\nabla w_\phi - \rho\mathbf{g} = 0,$$

where **g** is the acceleration of gravity.

The impermeability condition requires the normal diffusive current and velocity to vanish at the solid wall

$$\partial_n w = 0, \quad \mathbf{u}\cdot\mathbf{n} = 0, \quad \text{on } \Gamma. \tag{2.14}$$

where $\Gamma$ denotes boundaries of the domain. To remove singularities near the solid wall, the GNBC introduces the following slip boundary condition the fluid-solid surface [1, 16],

$$\beta\mathbf{u}_s = -\eta\partial_n\mathbf{u}_\tau + \lambda L(\phi)\nabla_\tau\phi, \quad \text{on } \Gamma, \tag{2.15}$$

where $\mathbf{u}_s = \mathbf{u}_\tau - \mathbf{u}_w$ is the slip velocity of fluid on the solid wall, $\mathbf{u}_w$ is the wall speed and $\mathbf{u}_\tau$ is the velocity along the boundary tangential direction $\tau$. $\nabla_\tau = \nabla - (\mathbf{n}\cdot\nabla)\mathbf{n}$ is the gradient along $\tau$. $\beta$ is the slip coefficient associated with the local composition at the solid surface, and it can be used to define a slip length $l_s = \eta/\beta$. For the sake of simplicity, we take $\beta$ as a constant in this work. Physically, $L(\phi)$ measures the deviation from equilibrium conditions at the solid wall, $-\eta\partial_n\mathbf{u}_\tau$ denotes the tangential viscous stress, and $\lambda L(\phi)\nabla_\tau\phi$ is the unbalanced Young stress [1]. The equation (2.15) indicates that the slip velocity at the solid wall is proportional to the sum of the tangential viscous stress and the unbalanced Young stress.

The GNBC also imposes the dynamic contact line condition at the solid wall

$$\phi_t + \mathbf{u}_\tau\cdot\nabla_\tau\phi = -\gamma L(\phi), \quad \text{on } \Gamma, \tag{2.16}$$

where $\gamma$ is a positive phenomenological relaxation parameter. The equation (2.16) means that the material derivative of $\phi$ at the solid surface is proportional to the deviation of $L$ from its equilibrium value [1]. In equilibrium, $L = 0$, the slip boundary condition (2.15) reduces to the Navier boundary condition $\beta\mathbf{u}_s = -\eta\partial_n\mathbf{u}_\tau$.

Equations (2.6) – (2.16) form a complete set of governing equations for the phase-field MCL model, and the total energy of the hydrodynamic system is the sum of kinetic energy $E_k$ and free energy $E_f$,

$$E_{tot}(\phi) = E_k + E_f = \int\left(\frac{\rho}{2}|\mathbf{u}|^2 + \frac{\lambda\varepsilon}{2}|\nabla\phi|^2 + \lambda F(\phi)\right)\mathrm{d}\Omega + \lambda\int M(\phi)\,\mathrm{d}\Gamma. \tag{2.17}$$

Assuming $\mathbf{u}_w = 0$, we can easily derive the following PDE energy dissipation law for the above governing system:

$$\frac{d}{dt}E_{tot} = -\frac{1}{2}\int\left|\sqrt{\eta}D(\mathbf{u})\right|^2\mathrm{d}\Omega - M_\phi\int|\nabla w|^2\mathrm{d}\Omega - \lambda\gamma\int|L(\phi)|^2\,\mathrm{d}\Gamma - \beta\int|\mathbf{u}_\tau|^2\,\mathrm{d}\Gamma \le 0. \tag{2.18}$$

## 3. Energy stable numerical scheme

### 3.1. Transformed governing system

The main challenge of solving the phase-field MCL model is to construct an accurate and energy stable numerical scheme. Several studies [30, 52] have demonstrated that numerical schemes may lead to spurious solutions if they do not satisfy discrete energy laws when time step sizes and mesh sizes are not carefully chosen. Hence, to accurately simulate the contact line dynamics, it is especially desirable to design schemes that satisfy the corresponding discrete energy laws. In this section, we present an accurate scheme with unconditional energy stability to solve the phase-field MCL model.

There are mainly four difficulties [16, 30] in developing the desired scheme, including (1) the treatment of nonlinear double well potential; (2) the nonlinear coupling terms between the phase-field variable and velocity through stress and convective terms; (3) the coupling of velocity and pressure through the incompressibility constraint; (4) the nonlinear term in $L(\phi)$ originating from the surface energy at fluid-solid surface.

For the difficulty (1), the commonly used techniques, such as convex splitting and stabilization approaches, may not be optimal choices due to some imperfections. The scheme constructed by the convex splitting approach usually involves a nonlinear coupled system, which is computationally expensive to obtain numerical results [42]. The stabilization approach requires the second derivative of the double well potential $F(\phi)$ to be bounded. Nevertheless, this is not satisfied by $F(\phi)$. Thus, a novel scalar auxiliary variable (SAV) approach is used in this work to deal with $F(\phi)$, and it has been successfully applied to solve a large class of gradient flows [44, 45]. For the difficulty (2), we apply some subtle explicit-implicit treatment to the nonlinear terms. For the difficulty (3), the pressure stabilization method [53] is used to decouple the computation of velocity and pressure, and it allows us to solve a Poisson equation with the constant coefficient. For the difficulty (4), a stabilization term is added artificially to balance the explicit nonlinear term [54].

The idea of SAV approach is natural and simple. A scalar auxiliary variable is introduced to transform the total energy into an equivalent form, and then the nonlinear term can be treated semi-explicitly [43, 44, 46]. More precisely, we define a new scalar variable

$$U=\sqrt{E_1(\phi)}=\sqrt{\int F(\phi)\,\mathrm{d}\Omega}, \tag{3.1}$$

and rewrite the total energy as

$$E_{tot}(\phi,U)=\int\left(\frac{\rho}{2}|\mathbf{u}|^2+\frac{\lambda\varepsilon}{2}|\nabla\phi|^2\right)\mathrm{d}\Omega+\lambda U^2+\lambda\int M(\phi)\,\mathrm{d}\Gamma, \tag{3.2}$$

Now a new and equivalent governing system can be obtained

$$\phi_t+\nabla\cdot(\mathbf{u}\phi)-M_\phi\Delta w=0, \tag{3.3}$$

$$w=-\lambda\varepsilon\Delta\phi+\frac{\lambda U}{\sqrt{E_1(\phi)}}F'(\phi), \tag{3.4}$$

$$U_t=\frac{1}{2\sqrt{E_1(\phi)}}\int_\Omega F'(\phi)\phi_t\,d\Omega, \tag{3.5}$$

$$\rho\mathbf{u}_t+\rho\mathbf{u}\cdot\nabla\mathbf{u}+\mathbf{J}\cdot\nabla\mathbf{u}-\nabla\cdot\eta D(\mathbf{u})+\nabla p+\phi\nabla w=0, \tag{3.6}$$

$$\nabla\cdot\mathbf{u}=0, \tag{3.7}$$

with the impermeability boundary condition at the solid wall

$$\partial_n w = 0, \quad \mathbf{u}\cdot\mathbf{n} = 0, \quad \text{on } \Gamma, \tag{3.8}$$

and the slip boundary condition

$$\beta \mathbf{u}_s = -\eta \partial_n \mathbf{u}_\tau + \lambda L(\phi)\nabla_\tau \phi, \quad \text{on } \Gamma, \tag{3.9}$$

as well as the dynamic contact line condition

$$\phi_t + \mathbf{u}_\tau \cdot \nabla_\tau \phi = -\gamma L(\phi), \quad \text{on } \Gamma, \tag{3.10}$$

where

$$L(\phi) = \varepsilon \partial_n \phi + M'(\phi), \quad \text{on } \Gamma. \tag{3.11}$$

Then we can derive the PDE energy dissipation law for the transformed system (3.3) – (3.11). Assuming $\mathbf{u}_w$ =0, then the transformed governing system satisfies the following energy dissipation law

$$\frac{d}{dt}E_{tot}(\phi, U) = -\frac{1}{2}\int\left|\sqrt{\eta}D(\mathbf{u})\right|^2 \mathrm{d}\Omega - M_\phi \int |\nabla w|^2 \mathrm{d}\Omega - \lambda\gamma\int\left|L(\phi)\right|^2 \mathrm{d}\Gamma - \beta\int|\mathbf{u}_\tau|^2 \mathrm{d}\Gamma \le 0.$$

We can see that the transformed system (3.3) – (3.11) satisfies the exactly same energy dissipation law with the original system for the time-continuous case. Next we will try to develop an energy stable time-marching scheme for the transformed system.

### 3.2. Energy stable time-marching scheme

We now present an energy stable scheme for the transformed governing system. To deal with the case of nonmatching density, a cut-off function is defined

$$\hat{\phi}^{n+1} = \begin{cases} \phi^{n+1} & |\phi^{n+1}| \le 1, \\ sign\left(\phi^{n+1}\right) & |\phi^{n+1}| > 1. \end{cases}$$

Given $\phi^n$, $\mathbf{u}^n$, $p^n$, $\rho^n$ and $\eta^n$, the scheme calculates $\phi^{n+1}$, $\mathbf{u}^{n+1}$, $p^{n+1}$, $\rho^{n+1}$ and $\eta^{n+1}$ in two steps. In the step 1, we solve the following coupled system to update $\phi^{n+1}$, $w^{n+1}$, $\mathbf{u}^{n+1}$, $\rho^{n+1}$ and $\eta^{n+1}$:

$$\frac{\phi^{n+1} - \phi^n}{\delta t} + \nabla\cdot\left(\mathbf{u}^{n+1}\phi^n\right) - M_\phi \Delta w^{n+1} = 0, \tag{3.12}$$

$$w^{n+1} = -\lambda\varepsilon\Delta\phi^{n+1} + \frac{\lambda U^{n+1}}{\sqrt{E_1\left(\phi^n\right)}} F'\left(\phi^n\right), \tag{3.13}$$

$$\frac{U^{n+1} - U^n}{\delta t} = \frac{1}{2\sqrt{E_1\left(\phi^n\right)}}\int_\Omega F'\left(\phi^n\right)\frac{\left(\phi^{n+1} - \phi^n\right)}{\delta t}\mathrm{d}\Omega \tag{3.14}$$

$$\begin{aligned} &\rho^n \frac{\mathbf{u}^{n+1} - \mathbf{u}^n}{\delta t} + \left(\rho^n \mathbf{u}^n \cdot \nabla\right)\mathbf{u}^{n+1} + \mathbf{J}^n \cdot \nabla \mathbf{u}^{n+1} - \nabla\cdot\eta^n D\left(\mathbf{u}^{n+1}\right) + \nabla\left(2p^n - p^{n-1}\right) + \\ &\phi^n \nabla w^{n+1} + \frac{1}{2}\frac{\rho^{n+1} - \rho^n}{\delta t}\mathbf{u}^{n+1} + \frac{1}{2}\nabla\cdot\left(\rho^n \mathbf{u}^n\right)\mathbf{u}^{n+1} + \frac{1}{2}\nabla\cdot\mathbf{J}^n \mathbf{u}^{n+1} = 0, \end{aligned} \tag{3.15}$$

where

$$\mathbf{J}^n = \frac{\rho_2 - \rho_1}{2} M_\phi \nabla w^n, \quad \rho^{n+1} = \frac{\rho_1 - \rho_2}{2}\hat{\phi}^{n+1} + \frac{\rho_1 + \rho_2}{2}, \quad \eta^{n+1} = \frac{\eta_1 - \eta_2}{2}\hat{\phi}^{n+1} + \frac{\eta_1 + \eta_2}{2},$$

with boundary conditions

$$\partial_n w^{n+1} = \mathbf{u}^{n+1} \cdot \mathbf{n} = 0, \quad \text{on } \Gamma, \tag{3.16}$$

$$\frac{\phi^{n+1} - \phi^n}{\delta t} + \mathbf{u}_\tau^{n+1} \cdot \nabla_\tau \phi^n = -\gamma \tilde{L}^{n+1}, \quad \text{on } \Gamma, \tag{3.17}$$

$$\beta \mathbf{u}_s^{n+1} + \eta^n \partial_n \mathbf{u}_\tau^{n+1} - \lambda \tilde{L}^{n+1} \nabla_\tau \phi^n = 0, \quad \text{on } \Gamma, \tag{3.18}$$

where

$$\tilde{L}^{n+1} = \varepsilon \partial_n \phi^{n+1} + M'\left(\phi^n\right) + S\left(\phi^{n+1} - \phi^n\right), \quad \text{on } \Gamma. \tag{3.19}$$

In the step 2, we update $p^{n+1}$. To avoid solving an elliptic equation with the variable coefficient $1/\rho$, the pressure-stabilized method is adopted to solve the pressure Poisson equation [30, 53]

$$\begin{cases} -\Delta\left(p^{n+1} - p^n\right) = -\dfrac{\chi}{\delta t} \nabla \cdot \mathbf{u}^{n+1}, \\ \partial_n p^{n+1} = 0, \quad \text{on } \Gamma, \end{cases} \tag{3.20}$$

where $\chi = \min\left(\rho_1, \rho_2\right)/2$.

**Remark 3.1.** (1) The introduction of the scalar variable $U$ allows the double well potential to be treated semi-implicitly. Once we obtain $\phi^{n+1}$, $U^{n+1}$ can be explicitly calculated by (3.14), and thus do not involve extra computational cost. Compared with the traditional approaches, e.g., the convex splitting approach, the SAV approach only requires the free energy term containing the nonlinear potential is bounded from below [43]. Also, this approach is not restricted to the specific forms of nonlinear parts of free energy, and it works well for the double well potential in this study. (2) A stabilization term $S\left(\phi^{n+1} - \phi^n\right)$ is added to balance the explicit surface energy potential $M'\left(\phi^n\right)$. $S$ is a positive constant with the magnitude determined later. In fact, the SAV approach can also be used to deal with the surface energy potential, but the resulting scheme will present a great challenge to the solution of the phase-field variable, and this is why we use the stabilization approach in this study. (3) The last three terms in (3.15) are a first-order approximation of the term $\left(\rho_t + \nabla \cdot \left(\rho \mathbf{u}\right) + \nabla \cdot \mathbf{J}\right)\mathbf{u}/2$. This term vanishes in (2.9) due to (2.13) [16, 30]. (4) Note that the nonlinearly coupled scheme (3.12) – (3.20) is an extension of our earlier work [46]. In [46], we can construct an energy-stable scheme, in which computations of the phase-field variable and velocity are decoupled by introducing an intermediate velocity. However, boundary conditions (3.9) – (3.11) introduced in the MCL model prohibit us from constructing a similar decoupled scheme.

Next we will rigorously prove that the scheme (3.12) – (3.20) is unconditionally energy stable. Compared to the phase-field two-phase flow model [46], the MCL model brings extra difficulties to the proof of energy stability. An effective strategy is needed to overcome these difficulties to obtain the desired energy dissipation law. Here we take a similar but different strategy to the reference [46] to prove the energy stability. The proof in this work starts with the difference between kinetic energies at $n$+1 and $n$ steps, but the proof in [46] starts from the Navier-Stokes equation. The treatment of the pressure equation is same. Before carrying out the energy estimate for the above scheme, we define a variable as follows

$$L_1 = \max\left|M''(\phi)\right| = \frac{\sqrt{2}\pi^2}{12}.$$

**Theorem 3.1.** Assuming $\mathbf{u}_w = 0$, and $S \geq L_1/2$, then the semi-implicit scheme (3.12) – (3.20) is energy stable and satisfies the following energy dissipation law:

$$E_{tot}^{n+1} - E_{tot}^{n} \leq -\frac{\delta t}{2}\left\|\sqrt{\eta^n} D\left(\mathbf{u}^{n+1}\right)\right\|^2 - \delta t M_\phi \left\|\nabla w^{n+1}\right\|^2 - \delta t \lambda \gamma \left\|\tilde{L}^{n+1}\right\|_\Gamma^2 - \delta t \beta \left\|\mathbf{u}_\tau^{n+1}\right\|_\Gamma^2, \tag{3.21}$$

where $\|\cdot\|$ denotes the $L^2$ - norm in $\Omega$ and

$$E_{tot}^{n} = \frac{1}{2}\left(\rho^n, \left|\mathbf{u}^n\right|^2\right) + \frac{\delta t^2}{2\chi}\left\|\nabla p^n\right\|^2 + \frac{\lambda\varepsilon}{2}\left\|\nabla \phi^n\right\|^2 + \lambda\left(U^n\right)^2 + \lambda\left(M\left(\phi^n\right), 1\right)_\Gamma. \tag{3.22}$$

**Proof.** The difference between kinetic energies $E_k^{n+1}$ and $E_k^n$ is estimated as [55, 56]

$$\begin{aligned}
E_k^{n+1} - E_k^n &= \frac{1}{2}\left(\rho^{n+1}, \left|\mathbf{u}^{n+1}\right|^2\right) - \frac{1}{2}\left(\rho^n, \left|\mathbf{u}^n\right|^2\right) \\
&= \frac{1}{2}\left(\rho^n, \left|\mathbf{u}^{n+1}\right|^2 - \left|\mathbf{u}^n\right|^2\right) + \frac{1}{2}\left(\rho^{n+1} - \rho^n, \left|\mathbf{u}^{n+1}\right|^2\right) \\
&= \left(\rho^n\left(\mathbf{u}^{n+1} - \mathbf{u}^n\right), \mathbf{u}^{n+1}\right) - \frac{1}{2}\left(\rho^n, \left|\mathbf{u}^{n+1} - \mathbf{u}^n\right|^2\right) + \frac{1}{2}\left(\rho^{n+1} - \rho^n, \left|\mathbf{u}^{n+1}\right|^2\right).
\end{aligned} \tag{3.23}$$

where $(\cdot, \cdot)$ denotes the inner product in $L^2(\Omega)$.
According to (3.15), we have

$$\begin{aligned}
\rho^n\left(\mathbf{u}^{n+1} - \mathbf{u}^n\right) = &-\delta t\left(\rho^n \mathbf{u}^n \cdot \nabla\right)\mathbf{u}^{n+1} - \delta t \mathbf{J}^n \cdot \nabla \mathbf{u}^{n+1} + \delta t \nabla \cdot \eta^n D\left(\mathbf{u}^{n+1}\right) - \delta t \nabla\left(2p^n - p^{n-1}\right) \\
&- \frac{\left(\rho^{n+1} - \rho^n\right)}{2}\mathbf{u}^{n+1} - \frac{\delta t}{2}\nabla \cdot \left(\rho^n \mathbf{u}^n\right)\mathbf{u}^{n+1} - \frac{\delta t}{2}\nabla \cdot \mathbf{J}^n \mathbf{u}^{n+1} - \delta t \phi^n \nabla w^{n+1}.
\end{aligned} \tag{3.24}$$

By taking the $L^2$ inner product of (3.24) with $\mathbf{u}^{n+1}$, and using (3.23) and the following identities

$$\left[\left(\rho^n \mathbf{u}^n \cdot \nabla\right)\mathbf{u}^{n+1} + \frac{1}{2}\nabla \cdot \left(\rho^n \mathbf{u}^n\right)\mathbf{u}^{n+1}, \mathbf{u}^{n+1}\right] = 0,$$

$$\left[\left(\mathbf{J}^n \cdot \nabla\right)\mathbf{u}^{n+1} + \frac{1}{2}\left(\nabla \cdot \mathbf{J}^n\right)\mathbf{u}^{n+1}, \mathbf{u}^{n+1}\right] = 0,$$

we can derive that

$$\begin{aligned}
E_k^{n+1} - E_k^n = &-\frac{\delta t}{2}\left\|\sqrt{\eta^n} D\left(\mathbf{u}^{n+1}\right)\right\|^2 - \delta t\left(p^{n+1} - 2p^n + p^{n-1}, \nabla \cdot \mathbf{u}^{n+1}\right) + \delta t\left(p^{n+1}, \nabla \cdot \mathbf{u}^{n+1}\right) \\
&- \frac{1}{2}\left(\rho^n, \left|\mathbf{u}^{n+1} - \mathbf{u}^n\right|^2\right) - \delta t\left(\phi^n \nabla w^{n+1}, \mathbf{u}^{n+1}\right) + \delta t\left(\eta^n \partial_n \mathbf{u}^{n+1}, \mathbf{u}^{n+1}\right)_\Gamma.
\end{aligned} \tag{3.25}$$

For the boundary term in (3.25), using (3.18), we have

$$\begin{aligned}
\delta t\left(\eta^n \partial_n \mathbf{u}^{n+1}, \mathbf{u}^{n+1}\right)_\Gamma &= \delta t\left(-\beta \mathbf{u}_s^{n+1} + \lambda \tilde{L}^{n+1} \nabla_\tau \phi^n, \mathbf{u}_\tau^{n+1}\right)_\Gamma \\
&= -\delta t \beta \left\|\mathbf{u}_\tau^{n+1}\right\|_\Gamma^2 + \delta t \lambda\left(\tilde{L}^{n+1} \nabla_\tau \phi^n, \mathbf{u}_\tau^{n+1}\right)_\Gamma.
\end{aligned} \tag{3.26}$$

Note that the slip velocity $\mathbf{u}_s$ is equal to the tangential velocity $\mathbf{u}_\tau$ due to the velocity of wall $\mathbf{u}_w$ being zero.

By taking the $L^2$ inner product of (3.20) with $\delta t^2(p^{n+1} - 2p^n + p^{n-1})/\chi$ and with $-\delta t^2 p^{n+1}/\chi$ separately, we obtain

$$
\begin{aligned}
&-\frac{\delta t^2}{2\chi}\left(\left\|\nabla\left(p^{n+1}-p^n\right)\right\|^2-\left\|\nabla\left(p^n-p^{n-1}\right)\right\|^2+\left\|\nabla\left(p^{n+1}-2p^n+p^{n-1}\right)\right\|^2\right)\\
&\qquad=\delta t\left(p^{n+1}-2p^n+p^{n-1},\nabla\cdot\mathbf{u}^{n+1}\right),
\end{aligned}
\tag{3.27}
$$

and

$$
\frac{\delta t^2}{2\chi}\left(\left\|\nabla p^{n+1}\right\|^2-\left\|\nabla p^n\right\|^2+\left\|\nabla\left(p^{n+1}-p^n\right)\right\|^2\right)=-\delta t\left(p^{n+1},\nabla\cdot\mathbf{u}^{n+1}\right).
\tag{3.28}
$$

Combining (3.27) and (3.28), we have

$$
\begin{aligned}
&-\delta t\left(p^{n+1}-2p^n+p^{n-1},\nabla\cdot\mathbf{u}^{n+1}\right)+\delta t\left(p^{n+1},\nabla\cdot\mathbf{u}^{n+1}\right)\\
&\quad=-\frac{\delta t^2}{2\chi}\left(\left\|\nabla p^{n+1}\right\|^2-\left\|\nabla p^n\right\|^2\right)-\frac{\delta t^2}{2\chi}\left\|\nabla\left(p^n-p^{n-1}\right)\right\|^2+\frac{\delta t^2}{2\chi}\left\|\nabla\left(p^{n+1}-2p^n+p^{n-1}\right)\right\|^2.
\end{aligned}
\tag{3.29}
$$

We then take the difference of (3.20) at step $t^{n+1}$ and $t^n$ to derive

$$
\frac{\delta t^2}{2\chi}\left\|\nabla\left(p^{n+1}-2p^n+p^{n-1}\right)\right\|^2\le\frac{\chi}{2}\left\|\mathbf{u}^{n+1}-\mathbf{u}^n\right\|^2\le\frac{1}{4}\left(\rho^n,\left|\mathbf{u}^{n+1}-\mathbf{u}^n\right|\right)^2.
\tag{3.30}
$$

Summing up equations (3.25), (3.26), (3.29) and (3.30), and drop off some positive terms, we can derive that

$$
\begin{aligned}
E_k^{n+1}-E_k^n\le&-\frac{\delta t}{2}\left\|\sqrt{\eta^n}D\left(\mathbf{u}^{n+1}\right)\right\|^2-\frac{\delta t^2}{2\chi}\left(\left\|\nabla p^{n+1}\right\|^2-\left\|\nabla p^n\right\|^2\right)\\
&-\delta t\left(\phi^n\nabla w^{n+1},\mathbf{u}^{n+1}\right)-\delta t\beta\left\|\mathbf{u}_\tau^{n+1}\right\|_\Gamma^2+\delta t\lambda\left(\tilde{L}^{n+1}\nabla_\tau\phi^n,\mathbf{u}_\tau^{n+1}\right)_\Gamma.
\end{aligned}
\tag{3.31}
$$

By taking the $L^2$ inner product of (3.12) with $\delta tw^{n+1}$, we have

$$
\left(\phi^{n+1}-\phi^n,w^{n+1}\right)-\delta t\left(\mathbf{u}^{n+1}\phi^n,\nabla w^{n+1}\right)+\delta tM_\phi\left\|\nabla w^{n+1}\right\|^2=0.
\tag{3.32}
$$

Taking the $L^2$ inner product of (3.13) with $-(\phi^{n+1}-\phi^n)$, we obtain

$$
\begin{aligned}
-\left(\phi^{n+1}-\phi^n,w^{n+1}\right)=&-\frac{\lambda\varepsilon}{2}\left(\left\|\nabla\phi^{n+1}\right\|^2-\left\|\nabla\phi^n\right\|^2+\left\|\nabla\phi^{n+1}-\nabla\phi^n\right\|^2\right)\\
&+\lambda\left(\varepsilon\partial_n\phi^{n+1},\phi^{n+1}-\phi^n\right)_\Gamma-\lambda\left(U^{n+1}b^n,\phi^{n+1}-\phi^n\right),
\end{aligned}
\tag{3.33}
$$

where $b^n=F'\left(\phi^n\right)\big/\sqrt{E_1\left(\phi^n\right)}$.

For the boundary term in (3.33), using equations (3.19), (3.17) and Taylor-expansion,

$$
M'\left(\phi^n\right)\left(\phi^{n+1}-\phi^n\right)=M\left(\phi^{n+1}\right)-M\left(\phi^n\right)-\frac{M''\left(\xi^n\right)}{2}\left(\phi^{n+1}-\phi^n\right)^2,
$$

we can derive

$$
\begin{aligned}
&\lambda\left(\varepsilon\partial_n\phi^{n+1},\phi^{n+1}-\phi^n\right)_\Gamma\\
&\quad=\delta t\lambda\left(\tilde{L}^{n+1},-\gamma\tilde{L}^{n+1}-\mathbf{u}_\tau^{n+1}\cdot\nabla_\tau\phi^n\right)_\Gamma-\lambda\left(M'\left(\phi^n\right)+S\left(\phi^{n+1}-\phi^n\right),\phi^{n+1}-\phi^n\right)_\Gamma\\
&\quad=-\delta t\lambda\gamma\left\|\tilde{L}\right\|_\Gamma^2-\delta t\lambda\left(\tilde{L}^{n+1}\nabla_\tau\phi^n,\mathbf{u}_\tau^{n+1}\right)_\Gamma-\lambda\left(M\left(\phi^{n+1}\right)-M\left(\phi^n\right),1\right)_\Gamma\\
&\quad\quad-\lambda\left(S-\frac{M''\left(\xi^n\right)}{2},\left(\phi^{n+1}-\phi^n\right)^2\right)_\Gamma.
\end{aligned}
\tag{3.34}
$$

Taking the $L^2$ inner product of (3.14) with $2\delta t\lambda U^{n+1}$, we get

$$
\lambda\left[\left(U^{n+1}\right)^2-\left(U^n\right)^2+\left(U^{n+1}-U^n\right)^2\right]=\lambda\left(U^{n+1}b^n,\phi^{n+1}-\phi^n\right).
\tag{3.35}
$$

Summing up equations (3.32) – (3.35), we derive that

$$
\begin{aligned}
&\frac{\lambda\varepsilon}{2}\left(\left\|\nabla\phi^{n+1}\right\|^2-\left\|\nabla\phi^{n}\right\|^2+\left\|\nabla\phi^{n+1}-\nabla\phi^{n}\right\|^2\right)+\lambda\left[\left(U^{n+1}\right)^2-\left(U^{n}\right)^2+\left(U^{n+1}-U^{n}\right)^2\right]\\
&\quad+\lambda\left(M\left(\phi^{n+1}\right)-M\left(\phi^{n}\right),1\right)_\Gamma\\
&\quad=\delta t\left(\mathbf{u}^{n+1},\phi^{n}\nabla w^{n+1}\right)-\delta t\lambda\left(\tilde{L}^{n+1}\nabla_\tau\phi^{n},\mathbf{u}_\tau^{n+1}\right)_\Gamma-\delta t M_\phi\left\|\nabla w^{n+1}\right\|^2-\delta t\lambda\gamma\left\|\tilde{L}\right\|_\Gamma^2\\
&\quad-\lambda\left(S-\frac{M''\left(\xi^n\right)}{2},\left(\phi^{n+1}-\phi^{n}\right)^2\right)_\Gamma .
\end{aligned}
\tag{3.36}
$$

Summing up equations (3.31) and (3.36), and dropping off some positive terms, we have

$$
\begin{aligned}
&\frac{1}{2}\left(\rho^{n+1},\left|\mathbf{u}^{n+1}\right|^2\right)-\frac{1}{2}\left(\rho^{n},\left|\mathbf{u}^{n}\right|^2\right)+\frac{\lambda\varepsilon}{2}\left(\left\|\nabla\phi^{n+1}\right\|^2-\left\|\nabla\phi^{n}\right\|^2\right)+\lambda\left[\left(U^{n+1}\right)^2-\left(U^{n}\right)^2\right]\\
&\quad+\frac{\delta t^2}{2\chi}\left(\left\|\nabla p^{n+1}\right\|^2-\left\|\nabla p^{n}\right\|^2\right)+\lambda\left(M\left(\phi^{n+1}\right)-M\left(\phi^{n}\right),1\right)_\Gamma\\
&\quad\le-\frac{\delta t}{2}\left\|\sqrt{\eta^n}D\left(\mathbf{u}^{n+1}\right)\right\|^2-\delta t M_\phi\left\|\nabla w^{n+1}\right\|^2-\delta t\beta\left\|\mathbf{u}_s^{n+1}\right\|_\Gamma^2\\
&\quad-\delta t\lambda\gamma\left\|\tilde{L}^{n+1}\right\|_\Gamma^2-\lambda\left(S-\frac{M''\left(\xi^n\right)}{2},\left(\phi^{n+1}-\phi^{n}\right)^2\right)_\Gamma .
\end{aligned}
\tag{3.37}
$$

By the assumption $S \geq L_1/2$, we get the desired result. □

**Remark 3.2.** (1) At the numerical level, the reformulated energy functional is different from the original energy functional because of the introduction of the term $\delta t^2\left\|\nabla p^n\right\|^2/2\chi$. Thus, the discrete energy dissipation law for the reformulated energy functional in (3.20) may not be available for the original energy functional. (2) Equations (3.12) – (3.20) form a nonlinearly coupled system. In our simulations, instead of using sub-iterations, we replace $\mathbf{u}^{n+1}$ by $\mathbf{u}^n$ in (3.12) and $\mathbf{u}_\tau^{n+1}$ by $\mathbf{u}_\tau^{n}$ in (3.17), then the computations of $\phi^{n+1}$, $\mathbf{u}^{n+1}$ and $p^{n+1}$ are totally decoupled. At each time step, we solve two linear equations with constant coefficients to obtain $\phi^{n+1}$. The solution procedure of $\phi^{n+1}$ can refer to [43]. The velocity $\mathbf{u}^{n+1}$ and the pressure $p^{n+1}$ can also be updated by solving linear elliptic equations. Thus, the simplified scheme is decoupled and efficient. However, the simplification will destroy the unconditional energy stability of the nonlinear scheme (3.12) – (3.20). Small time steps are needed to obtain the desired accuracy and energy stability since the decoupled scheme is conditionally energy sable.

## 4. Spatial discretization and energy stability analysis

A finite difference method on the staggered grids is implemented for the spatial discretization. The 2D computational domain is $\Omega = [0, L_x] \times [0, L_y]$, where $L_x$ and $L_y$ are positive real numbers. The domain $\Omega$ is divided into rectangular meshes, and mesh centers [57, 58] are located at

$$
x_i=\left(i-\frac{1}{2}\right)h_x,\quad i=1,\cdots,n_x,\qquad y_j=\left(j-\frac{1}{2}\right)h_y,\quad j=1,\cdots,n_y,
$$

where $h_x = L_x/n_x$ and $h_y = L_y/n_y$ are mesh sizes, and $n_x$ and $n_y$ are the number of meshes in each direction. Then we define four sets of uniform grid points as follows

$$E^{ew}=\left\{\left(x_{i-1/2},\,y_j\right)\middle|\, i=1,\cdots,n_x+1;\ \ j=0,\cdots,n_y+1\right\},\quad E^{ns}=\left\{\left(x_i,\,y_{j-1/2}\right)\middle|\, i=0,\cdots,n_x+1;\ \ j=1,\cdots,n_y+1\right\},$$

$$C=\left\{\left(x_i,\,y_j\right)\middle|\, i=0,\cdots,n_x+1;\ \ j=0,\cdots,n_y+1\right\},\qquad V=\left\{\left(x_{i-1/2},\,y_{j-1/2}\right)\middle|\, i=1,\cdots,n_x+1;\ \ j=1,\cdots,n_y+1\right\},$$

where $E^{ew}$ is the set of west-east edge points, $E^{ns}$ is the set of south-north edge points, $C$ is the set of cell-centered points and $V$ is the set of vertex-centered points. Points of ghost cells are also included in above point sets. We continue to define the following function spaces

$$U_h=\left\{u:E^{ew}\rightarrow\mathbb{R}\right\},\quad V_h=\left\{v:E^{ns}\rightarrow\mathbb{R}\right\},\quad P_h=\{\phi:C\rightarrow\mathbb{R}\},\quad N_h=\{\psi:V\rightarrow\mathbb{R}\}.$$

The horizontal component $u$ of velocity $\mathbf{u}$ is defined on $U_h$, and the vertical component $v$ of $\mathbf{u}$ is defined on $V_h$. $\phi$, $w$, $p$, $\rho$ and $\eta$ are defined on $P_h$.

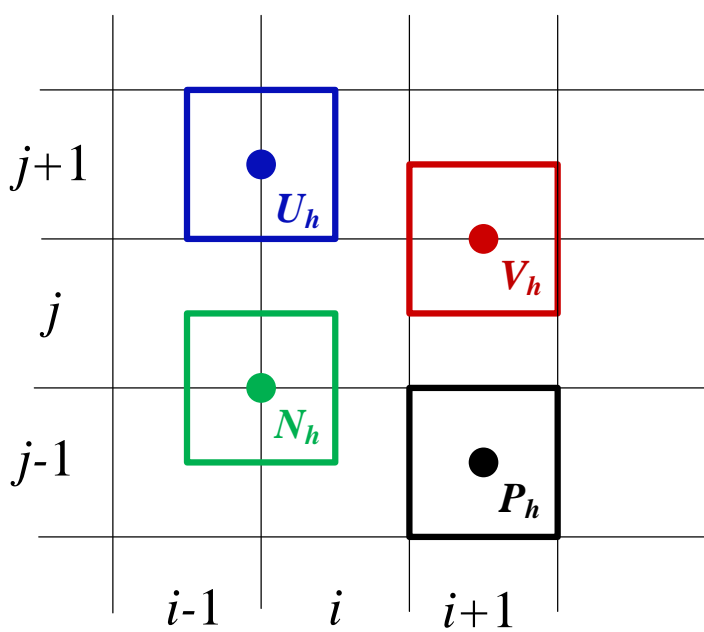

**Figure 1**. Finite difference spaces $U_h$, $V_h$, $P_h$ and $N_h$

Some finite difference operators [57-59] are defined to complete the spatial discretization:

(a) west-east-edge-to-center difference operator $d_x$: $U_h\longrightarrow P_h$ & $N_h\longrightarrow V_h$

$$d_x u_{i,j}=\frac{1}{h_x}\left(u_{i+1/2,j}-u_{i-1/2,j}\right),\quad d_x\psi_{i,j+1/2}=\frac{1}{h_x}\left(\psi_{i+1/2,j+1/2}-\psi_{i-1/2,j+1/2}\right);$$

(b) south-north-edge-to-center difference operator $d_y$: $V_h\longrightarrow P_h$ & $N_h\longrightarrow U_h$

$$d_y v_{i,j}=\frac{1}{h_y}\left(v_{i,j+1/2}-v_{i,j-1/2}\right),\quad d_y\psi_{i+1/2,j}=\frac{1}{h_y}\left(\psi_{i+1/2,j+1/2}-\psi_{i+1/2,j-1/2}\right);$$

(c) center-to-west-east-edge difference operator $D_x$: $P_h\longrightarrow U_h$ & $V_h\longrightarrow N_h$

$$D_x\phi_{i+1/2,j}=\frac{1}{h_x}\left(\phi_{i+1,j}-\phi_{i,j}\right),\quad D_x\psi_{i+1/2,j+1/2}=\frac{1}{h_x}\left(\psi_{i+1,j+1/2}-\psi_{i,j+1/2}\right);$$

(d) center-to-south-north-edge difference operator $D_y$: $P_h\longrightarrow V_h$ & $U_h\longrightarrow N_h$

$$D_y\phi_{i,j+1/2}=\frac{1}{h_y}\left(\phi_{i,j+1}-\phi_{i,j}\right),\quad D_y\psi_{i+1/2,j+1/2}=\frac{1}{h_y}\left(\psi_{i+1/2,j+1}-\psi_{i+1/2,j}\right);$$

(e) west-east average operator $A_x$: $P_h\longrightarrow U_h$, $U_h\longrightarrow P_h$ & $V_h\longrightarrow N_h$

$$A_x\phi_{i+1/2,j}=\frac{1}{2}\left(\phi_{i+1,j}+\phi_{i,j}\right),\quad A_x u_{i,j}=\frac{1}{2}\left(u_{i+1/2,j}+u_{i-1/2,j}\right),\quad A_x\psi_{i+1/2,j+1/2}=\frac{1}{2}\left(\psi_{i+1,j+1/2}+\psi_{i,j+1/2}\right);$$

(f) south-north average operator $A_y$: $P_h\longrightarrow V_h$, $V_h\longrightarrow P_h$ & $U_h\longrightarrow N_h$

$$A_y\phi_{i,j+1/2}=\frac{1}{2}\left(\phi_{i,j+1}+\phi_{i,j}\right),\quad A_y v_{i,j}=\frac{1}{2}\left(v_{i,j+1/2}+v_{i,j-1/2}\right),\quad A_y\psi_{i+1/2,j+1/2}=\frac{1}{2}\left(\psi_{i+1/2,j+1}+\psi_{i+1/2,j}\right).$$

With these finite difference operators, we can fully discretize the scheme (3.12) – (3.20) in a 2D domain, and the 3D spatial discretization is similar.

$$\phi^{n+1}-\phi^{n}+\delta t\left(d_x\left(u^{n+1}A_x\phi^{n}\right)+d_y\left(v^{n+1}A_y\phi^{n}\right)\right)-\delta t M_\phi\left(d_x\left(D_x w^{n+1}\right)+d_y\left(D_y w^{n+1}\right)\right)=0,\qquad(4.1)$$

$$w^{n+1} = -\lambda\varepsilon\left(d_x\left(D_x\phi^{n+1}\right) + d_y\left(D_y\phi^{n+1}\right)\right) + \lambda U^{n+1} b^n, \tag{4.2}$$

$$U^{n+1} - U^n = \frac{h_x h_y}{2}\sum_{i,j} b_{i,j}^n\left(\phi_{i,j}^{n+1} - \phi_{i,j}^n\right), \tag{4.3}$$

where $b^n = F'\left(\phi_{i,j}^n\right) \Big/ \sqrt{h_x h_y \sum_{i,j} F\left(\phi_{i,j}^n\right)}$.

$$\begin{aligned}
&A_x\rho^n\left(u^{n+1} - u^n\right) - \delta t\left(2D_x\left(\eta d_x u^{n+1}\right) + d_y\left(A_y\left(A_x\eta^n\right)D_y u^{n+1}\right) + d_y\left(A_y\left(A_x\eta^n\right)D_x v^{n+1}\right)\right)\\
&\quad + \delta t D_x\left(2p^n - p^{n-1}\right) + \delta t A_x\rho^n\left(u^n D_x\left(A_x u^{n+1}\right) + A_x\left(A_y v^n\right)d_y\left(A_y u^{n+1}\right)\right) + \delta t A_x\phi^n D_x w^{n+1}\\
&\quad + \delta t\left(J_x^n D_x\left(A_x u^{n+1}\right) + A_x\left(A_y J_y^n\right)d_y\left(A_y u^{n+1}\right)\right) + \frac{1}{2}u^{n+1}\left(A_x\rho^{n+1} - A_x\rho^n\right)\\
&\quad + \frac{1}{2}\delta t\left(D_x\left(A_x J_x^n\right) + d_y\left(A_x J_y^n\right)\right) + \frac{1}{2}\delta t A_x\rho^n\left(D_x\left(A_x u^n\right) + d_y\left(A_x v^n\right)\right)u^{n+1} = 0,
\end{aligned} \tag{4.4}$$

$$\begin{aligned}
&A_y\rho^n\left(v^{n+1} - v^n\right) - \delta t\left(d_x\left(A_x\left(A_y\eta^n\right)D_y u^{n+1}\right) + d_x\left(A_x\left(A_y\eta^n\right)D_x v^{n+1}\right) + 2D_y\left(\eta^n d_y v^{n+1}\right)\right)\\
&\quad + \delta t D_y\left(2p^n - p^{n-1}\right) + \delta t A_y\rho^n\left(A_x\left(A_y u^n\right)d_x\left(A_x v^{n+1}\right) + v^n D_y\left(A_y v^{n+1}\right)\right) + \delta t A_y\phi^n D_y w^{n+1}\\
&\quad + \delta t\left(A_y\left(A_x J_x^n\right)d_x\left(A_x v^{n+1}\right) + J_y^n D_y\left(A_y v^{n+1}\right)\right) + \frac{1}{2}v^{n+1}\left(A_y\rho^{n+1} - A_y\rho^n\right)\\
&\quad + \frac{1}{2}\delta t\left(d_x\left(A_y J_x^n\right) + D_y\left(A_y J_y^n\right)\right) + \frac{1}{2}\delta t A_y\rho^n\left(d_x\left(A_y u^n\right) + D_y\left(A_y v^n\right)\right)v^{n+1} = 0,
\end{aligned} \tag{4.5}$$

$$d_x\left(D_x\left(p^{n+1} - p^n\right)\right) + d_y\left(D_y\left(p^{n+1} - p^n\right)\right) = \frac{\chi}{\delta t}\left(d_x u^{n+1} + d_y v^{n+1}\right). \tag{4.6}$$

East and west boundaries of the rectangular domain are denoted by $\Gamma_{ew}$, and the remaining boundaries are denoted by $\Gamma_{ns}$. An operator $\mathbf{D}$ = ($D_x$, $D_y$) is introduced to simplify the discretization. The fully discrete boundary conditions are given based on these operators:

$$\begin{cases}
\mathbf{D}w^{n+1}\cdot\mathbf{n} = 0, \quad \text{on } \Gamma,\\
\phi^{n+1} - \phi^n + \delta t A_x\left(A_y u^{n+1}\right)d_x\left(A_x\left(A_y\phi^n\right)\right) + \gamma\tilde{L}^{n+1} = 0, \quad \text{on } \Gamma_{ns},\\
\phi^{n+1} - \phi^n + \delta t A_y\left(A_x v^{n+1}\right)d_y\left(A_y\left(A_x\phi^n\right)\right) + \gamma\tilde{L}^{n+1} = 0, \quad \text{on } \Gamma_{ew},
\end{cases} \tag{4.7}$$

$$\begin{cases}
\mathbf{u}^{n+1}\cdot\mathbf{n} = 0, \quad \text{on } \Gamma,\\
\beta A_y u^{n+1} + \sigma_{ns}A_y\left(A_x\eta^n\right)D_y u^{n+1} - \lambda A_x\tilde{L}^{n+1}D_x\left(A_y\phi^n\right) = 0, \quad \text{on } \Gamma_{ns},\\
\beta A_x v^{n+1} + \sigma_{ew}A_x\left(A_y\eta^n\right)D_x v^{n+1} - \lambda A_y\tilde{L}^{n+1}D_y\left(A_x\phi^n\right) = 0, \quad \text{on } \Gamma_{ew},\\
\mathbf{D}p^{n+1}\cdot\mathbf{n} = 0, \quad \text{on } \Gamma,
\end{cases} \tag{4.8}$$

where

$$\begin{cases}
\tilde{L}^{n+1} - M'\left(\phi^n\right) - S\left(\phi^{n+1} - \phi^n\right) - \sigma_{ns}\varepsilon D_y\phi^{n+1} = 0, \quad \text{on } \Gamma_{ns},\\
\tilde{L}^{n+1} - M'\left(\phi^n\right) - S\left(\phi^{n+1} - \phi^n\right) - \sigma_{ew}\varepsilon D_x\phi^{n+1} = 0, \quad \text{on } \Gamma_{ew},
\end{cases} \tag{4.9}$$

$$\rho^{n+1} = \frac{\rho_1 - \rho_2}{2}\hat{\phi}^{n+1} + \frac{\rho_1 + \rho_2}{2}, \quad \eta^{n+1} = \frac{\eta_1 - \eta_2}{2}\hat{\phi}^{n+1} + \frac{\eta_1 + \eta_2}{2},$$

$$J_x^n = \frac{\left(\rho_2 - \rho_1\right)}{2}M_\phi D_x w^n, \quad J_y^n = \frac{\left(\rho_2 - \rho_1\right)}{2}M_\phi D_y w^n,$$

$$\sigma_{ns} = \begin{cases} 1, & \text{on } \Gamma_n \\ -1, & \text{on } \Gamma_s \end{cases}, \quad \sigma_{ew} = \begin{cases} 1, & \text{on } \Gamma_e \\ -1, & \text{on } \Gamma_w \end{cases},$$

where $J_x$ is defined on $U_h$, and $J_y$ is defined on $V_h$.

**Remark 3.3.** In the above fully discrete scheme, we use the central difference scheme to discretize advection terms in the Cahn–Hilliard and Navier–Stokes equations. It is well known that the central difference scheme is inaccurate and unstable at large Péclet number, where the advection dominates the fluid flow. Hence, in all our simulations, advection terms are discretized by a composite high resolution scheme. More precisely, the fluxes at cell faces are evaluated with a MINMOD scheme [60, 61], which not only achieves the second-order accuracy in space, but also preserves the physical properties of convection [61]. A preconditioned biconjugate gradient stabilized method (BICGSTAB) is used to solve the above variables [62-64].

To simplify the proof of energy stability, we define some weighted inner products (see [57, 58]):

$$\langle \phi, \varphi \rangle_{P_h} = \sum_{i=1}^{nx} \sum_{j=1}^{ny} \phi_{i,j} \varphi_{i,j}, \qquad \phi, \varphi \in P_h, \tag{4.10}$$

$$\langle f, g \rangle_{U_h} = \frac{1}{2} \sum_{i=1}^{nx} \sum_{j=1}^{ny} \left( f_{i+1/2,\,j} g_{i+1/2,\,j} + f_{i-1/2,\,j} g_{i-1/2,\,j} \right), \qquad f, g \in U_h, \tag{4.11}$$

$$\langle f, g \rangle_{V_h} = \frac{1}{2} \sum_{i=1}^{nx} \sum_{j=1}^{ny} \left( f_{i,\,j+1/2} g_{i,\,j+1/2} + f_{i,\,j-1/2} g_{i,\,j-1/2} \right), \qquad f, g \in V_h. \tag{4.12}$$

The following weighted inner products are also defined to deal with boundary conditions. If $f$ and $g \in U_h$, then

$$\langle f, g \rangle_{\Gamma_{ew}} = -\left(1/h_x\right) \sum_{j=1}^{ny} g_{1/2,\,j} f_{1/2,\,j} + \left(1/h_x\right) \sum_{j=1}^{ny} g_{nx+1/2,\,j} f_{nx+1/2,\,j}. \tag{4.13}$$

If $f$ and $g \in V_h$, then

$$\langle f, g \rangle_{\Gamma_{ns}} = -\left(1/h_y\right) \sum_{i=1}^{nx} g_{i,\,1/2} f_{i,\,1/2} + \left(1/h_y\right) \sum_{i=1}^{nx} g_{i,\,ny+1/2} f_{i,\,ny+1/2}. \tag{4.14}$$

For any cell-centered function $\phi$ defined on $P_h$, we define

$$\|\phi\|_2 = \sqrt{h_x h_y \langle \phi, \phi \rangle_{P_h}},$$

$$\|\nabla \phi\|_2 = \sqrt{h_x h_y \langle D_x \phi, D_x \phi \rangle_{U_h} + h_x h_y \langle D_y \phi, D_y \phi \rangle_{V_h}}.$$

For the velocity $\mathbf{u}$, we define

$$\|\mathbf{u}\|_2 = \sqrt{h_x h_y \langle u, u \rangle_{U_h} + h_x h_y \langle v, v \rangle_{V_h}}.$$

**Theorem 2.2.** Assuming $\mathbf{u}_w = 0$, and $S \geq L_1/2$, then the fully discrete scheme (4.1) – (4.9) is energy stable and satisfies the following discrete energy law:

$$\frac{E_{tot}^{n+1} - E_{tot}^{n}}{\delta t} \leq -R_v^n - R_d^n - R_s^n - R_r^n, \tag{4.15}$$

where

$$R_v^n = \frac{1}{2}\left\| \sqrt{\eta^n} D\left(\mathbf{u}^{n+1}\right) \right\|_2^2, \quad R_d^n = M_\phi \left\| \nabla w^{n+1} \right\|_2^2, \quad R_s^n = \beta \left\| \mathbf{u}_\tau^{n+1} \right\|_{\Gamma, 2}^2, \quad R_r^n = \lambda \gamma \left\| \tilde{L}^{n+1} \right\|_{\Gamma, 2}^2 \tag{4.16}$$

$$E_{tot}^n = \frac{1}{2}\left\| \sqrt{\rho^n} \mathbf{u}^n \right\|_2^2 + \frac{\lambda \varepsilon}{2} \left\| \nabla \phi^n \right\|_2^2 + \lambda \left(U^n\right)^2 + \lambda \left(M\left(\phi^n\right), 1\right)_{\Gamma, 2} + \frac{\delta t^2}{2\chi} \left\| \nabla p^n \right\|_2^2, \tag{4.17}$$

here

$$U^n = \sqrt{h_x h_y \langle F\left(\phi^n\right), 1 \rangle_{P_h}}, \quad \left(M\left(\phi^n\right), 1\right)_{\Gamma, 2} = h_x h_y \langle \sigma_{ns} M\left(\phi^n\right), 1 \rangle_{\Gamma_{NS}} + h_x h_y \langle \sigma_{ew} M\left(\phi^n\right), 1 \rangle_{\Gamma_{EW}}.$$

$E_{tot}^n$ is the total energy, $R_v^n$ is the viscous dissipation rate, $R_d^n$ is the diffusion dissipation rate, $R_s^n$ is the dissipation rate induced by the fluid slip at the solid wall, and $R_r^n$ is the dissipation rate associated with composition relaxation at the fluid-solid interface [1].

**Proof.** The detailed derivation can refer to the Appendix A.

**Remark 3.4.** The SAV and stabilization approaches allow us to construct a fully discrete energy stable scheme, but they will definitely cause some errors. Therefore, it is necessary to conduct error and convergence analysis for the proposed scheme. Recently, the authors in [65] performed error and convergence analysis for the fully discrete Cahn-Hilliard phase-field model based on the SAV approach, and they further extended the analysis to the hydrodynamics coupled phase-field model [66]. In [67], the authors used the stabilization approach to deal with the nonlinear double well potential and conducted the error analysis. The above studies demonstrate that both SAV and stabilization approaches lead to a convergence system, but we still cannot theoretically guarantee the convergence of a numerical scheme constructed by the combined SAV and stabilization approach. It is quite a challenging task to conduct the convergence analysis of the fully discrete scheme for the phase-field MCL model, and this is what we are going to do next. However, some numerical experiments in Section 5 demonstrate that our results can converge to unique solutions.

## 5. Numerical Results

In this section, several numerical experiments are conducted to validate accuracy and energy stability of the fully discrete scheme.

### 5.1. Accuracy test

We first use the case of Couette flow [68] to investigate the accuracy and convergence rates for proposed scheme. All simulations are conducted in a 2D domain Ω = [0, 3] × [0, 1] with periodic boundary conditions applied on the left and right boundaries. The top and bottom walls move at the equal but opposite velocities $u_w^{\pm} = 0.2$ and the GNBCs are imposed on them. The initial profile of $\phi$ is set as

$$\phi\left(x, y\right) = \tanh\left( \frac{1}{\sqrt{2}\varepsilon} \left( \frac{L_x}{4} - \left| x - \frac{L_x}{2} \right| \right) \right), \tag{5.1}$$

where $L_x$ is the width of domain and other parameters used in simulations are as follows:

$\rho_1 = 1$, $\rho_2 = 0.9$, $\eta_1 = 1$, $\eta_2 = 1.1$, $M_\phi = 1 \times 10^{-3}$, $\varepsilon = 0.01$, $\lambda = 1.2$, $\gamma = 100$, $\theta_s = 60°$, $\beta = 5.26$.

Note that parameters used in this work mainly come from [16] and [41].

To assess the order of accuracy, we calculate the relative $L^{\infty}$ norm errors [69] for the horizontal velocity $u$, the pressure $p$ and the phase-field variable $\phi$. A spatial resolution of $N\times N/3$ ($N$ = 150, 300, 450, 600 and 750) is used to discretize the computational domain and the time step-size is taken as $\delta t$ = 0.1 $h$, where $h$ is the minimum grid size. We run each simulation until $t$ =3 (Figure 2). The solution at $N$ = 750 is used as a reference solution. The relative $L^{\infty}$ norm errors at time $t$ =3 are calculated, and results are shown in Figure 3. It can be observed that our scheme can achieve the second-order accuracy in space.

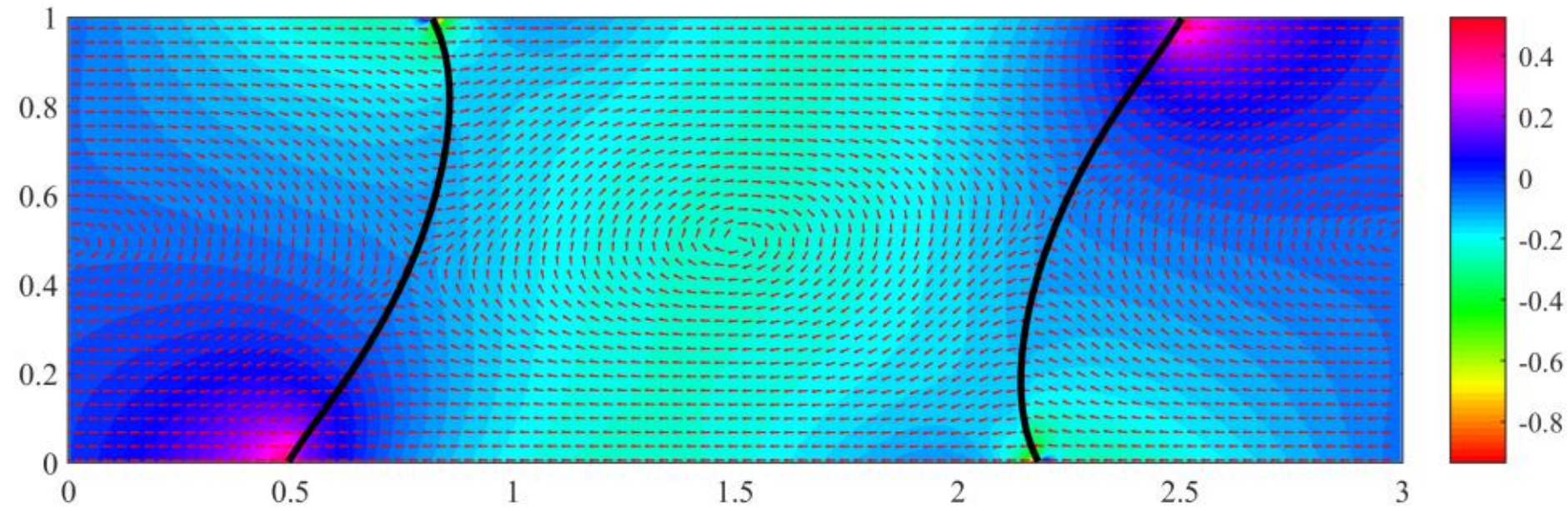

**Figure 2**. The pressure field (background color), contour line of $\phi$ = 0 and the quiver plot of velocity $u$, $v$ at $t$ =3.

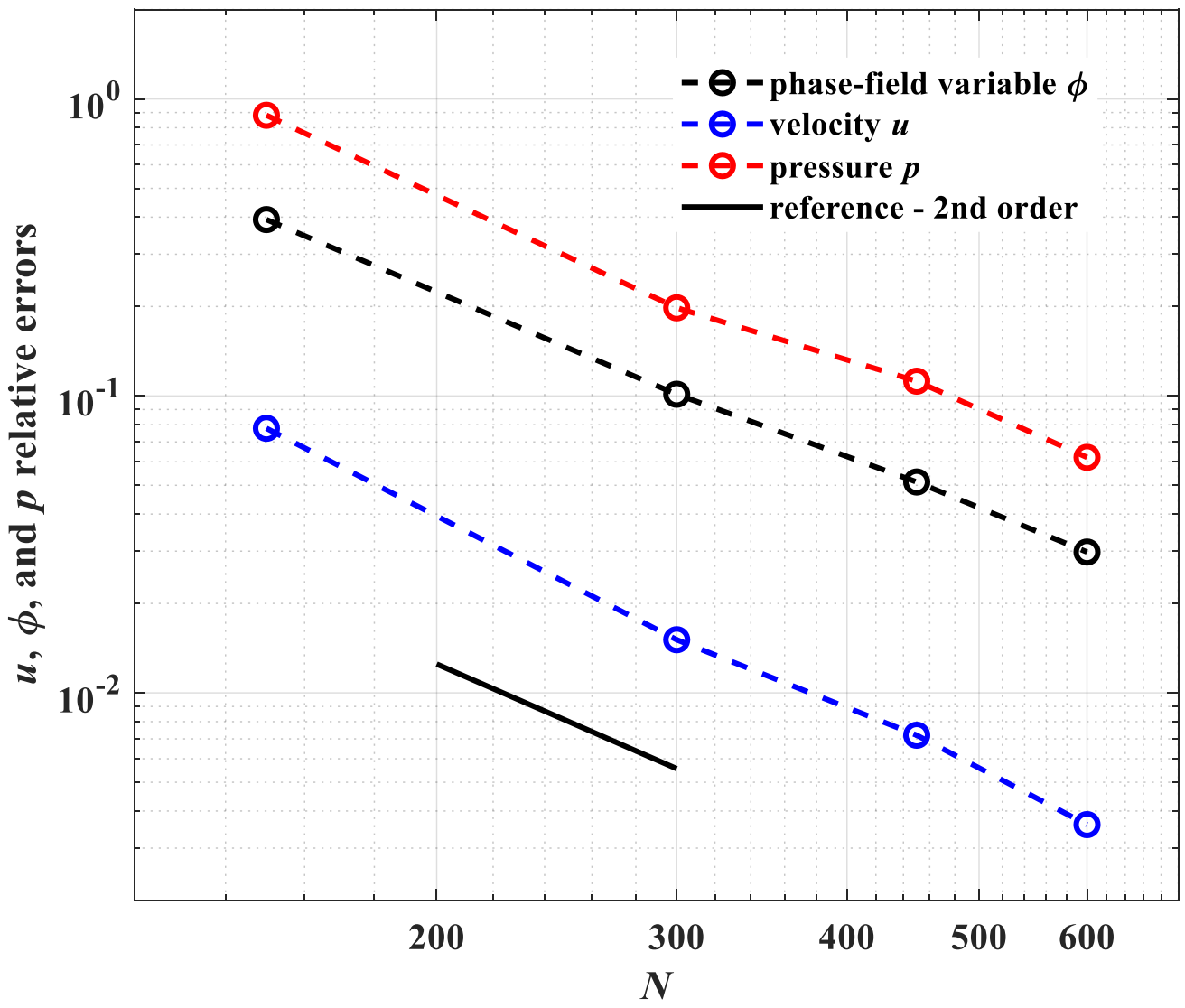

Figure 3. Relative $L^{\infty}$ errors as a function of $N$ for the velocity $u$ (blue), the pressure $p$ (black) and the phase-field variable $\phi$ (red).

Now, we verify the discrete energy law of the fully discrete scheme. To ensure no energy is input, we set the velocities of walls $u_w^{\pm} = 0$. The spatial resolution is 300 $\times$ 100, and the interfacial thickness parameter $\varepsilon$ is 0.02. We test the energy evolution from $\delta t$ = 0.02. Figure 4 shows the evolutions of the total energy at five different time step-sizes. The tendency of energy curves confirms that our scheme is energy stable. We also observe obvious differences between energies obtained by different time step-sizes, which indicate that large time step-sizes induce more numerical errors.

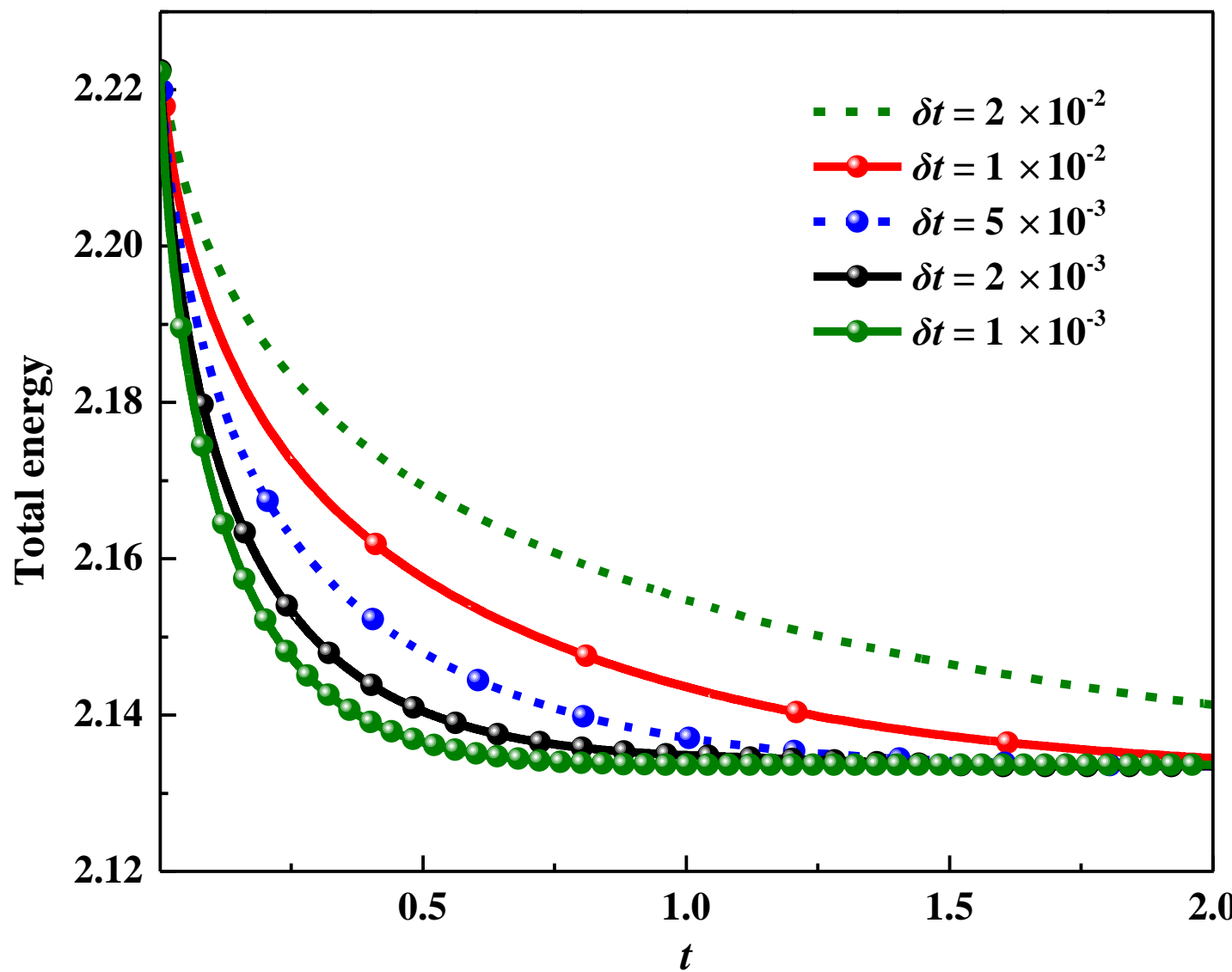

**Figure 4**. Evolution of the total energy $E_{tot}$ at different time step-sizes.

In addition to the SAV approach, some other approaches are also proposed to construct numerical schemes for the phase-field MCL model, such as the convex splitting and IEQ approaches. Here we only compare SAV and IEQ approaches using the Couette flow case. With reference to [70] and [71], we can also easily construct a nonlinearly coupled scheme for the variable density case based on the IEQ approach, and such a scheme can be simplified into a decoupled scheme by the same explicit treatment to the velocity as in Remark 3.2. Two different spatial resolutions ($n_x \times n_y = 300 \times 100$ and $450 \times 150$) are considered in this study, and the time step-size is $\delta t = 0.1\ h$. We run each simulation until $t = 3$.

**Table 1.** Computational time of SAV and IEQ approaches at different spatial resolutions

| | 300 × 100 | 450 × 150 |
|---|---|---|
| SAV | 779 s | 3514 s |
| IEQ | 885 s | 3990 s |

Table 1 gives the computational cost of two approaches at different spatial resolutions. Obviously, the decoupled SAV scheme is more efficient than the decoupled IEQ scheme. At each time step, the decoupled SAV scheme needs to solve two linear elliptic equations with constant coefficients to update $\phi^{n+1}$, so such a scheme is efficient, while the decoupled IEQ scheme needs to solve a linear equation with the variable coefficient. Figure 5 gives evolutions of left contact points. It can be observed that the curve (green line) obtained by the decoupled IEQ scheme at the spatial resolution 450 × 150 is consistent with the result (green points) by the decoupled SAV scheme at the spatial resolution 300 × 100. This result indicates that the SAV approach may be more accurate than the IEQ approach.

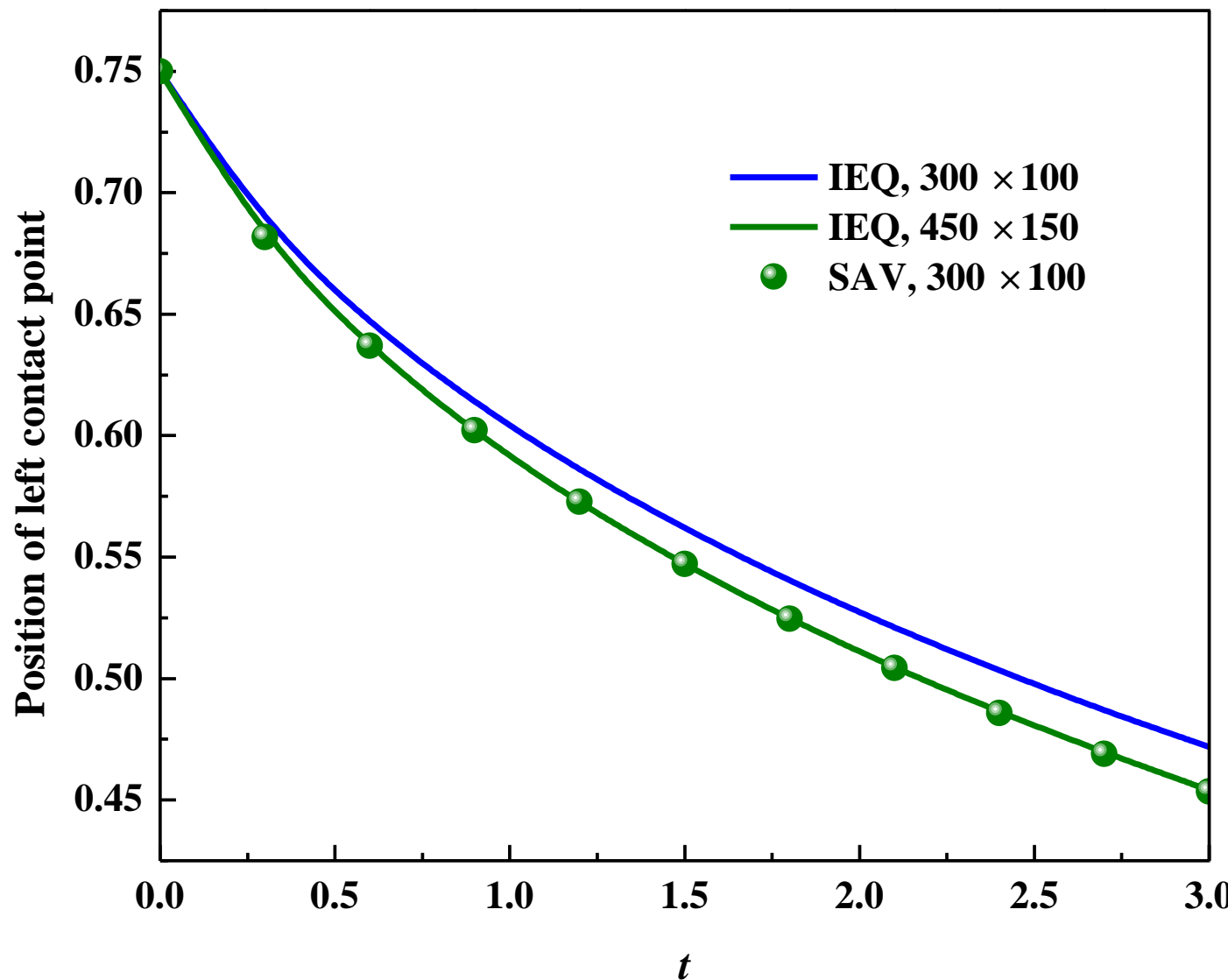

**Figure 5.** Positions of left contact points obtained by the IEQ (blue line: $n_x \times n_y = 300 \times 100$; green line: $450 \times 150$) and SAV approaches (green points: $300 \times 100$).

We next use the Couette flow case to contact analysis of the sharp interface limit for the MCL [71-73]. The top and bottom walls move at equal but opposite velocities 0.1. We consider four different $\varepsilon$ values ($\varepsilon$ = 0.04, 0.02, 0.01 and 0.006) while all other parameters are kept constant. A grid size $600 \times 200$ and time step-size $\delta t = 0.1\ h_x$ in all simulations. We run each simulation until $t$ = 10. Figure 6 gives interfacial shapes at $t$ =10 for four different $\varepsilon$ values, and Figure 7 presents evolutions of left contact points during the whole process. Numerical results demonstrate that the interface converges to a unique solution when the interfacial thickness parameter $\varepsilon$ is reduced. It seems reasonable to take $\varepsilon = 0.01$ to be the threshold for convergence to the sharp interface limit.

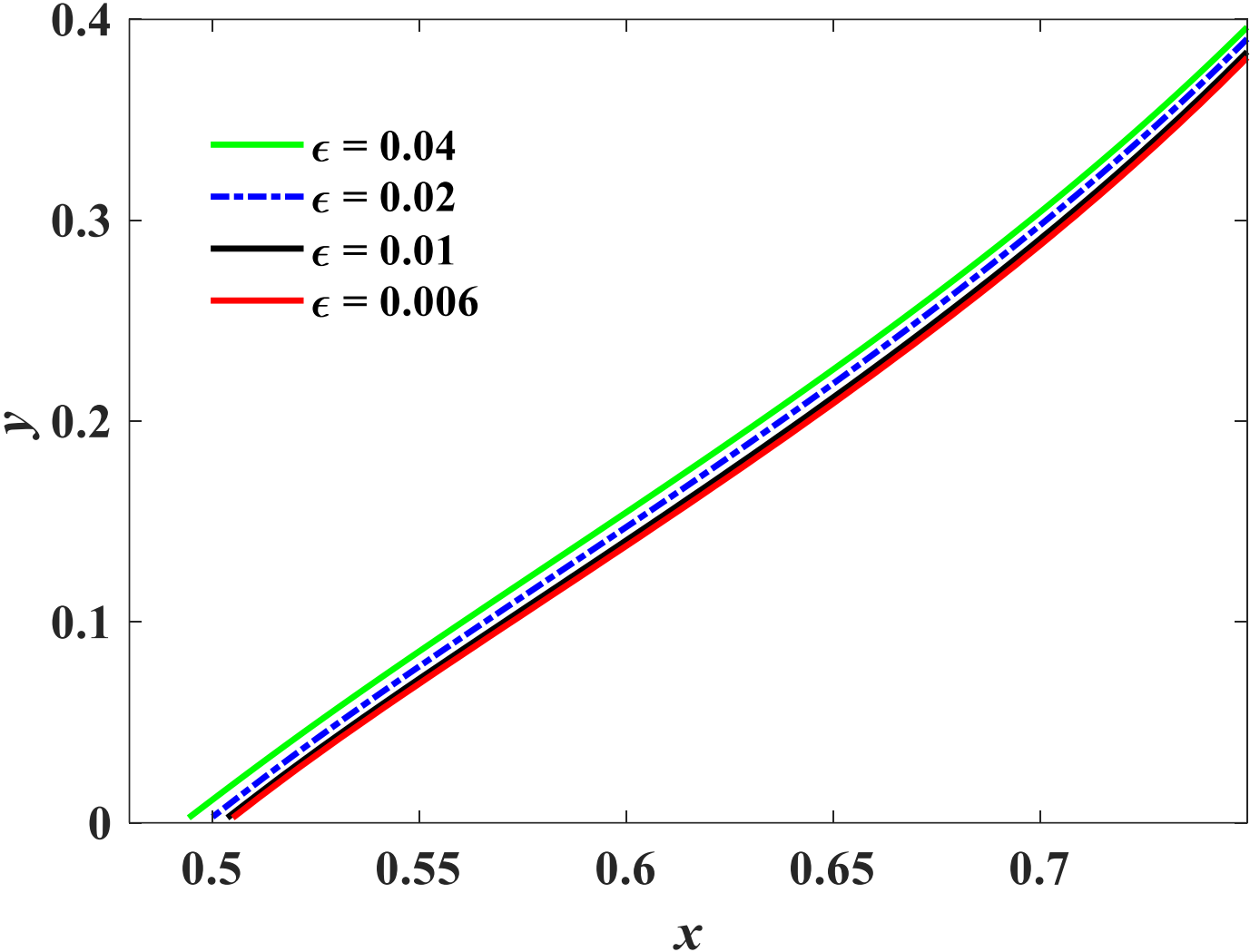

**Figure 6**. Interfacial shapes in Couette flow for four different $\varepsilon$ values.

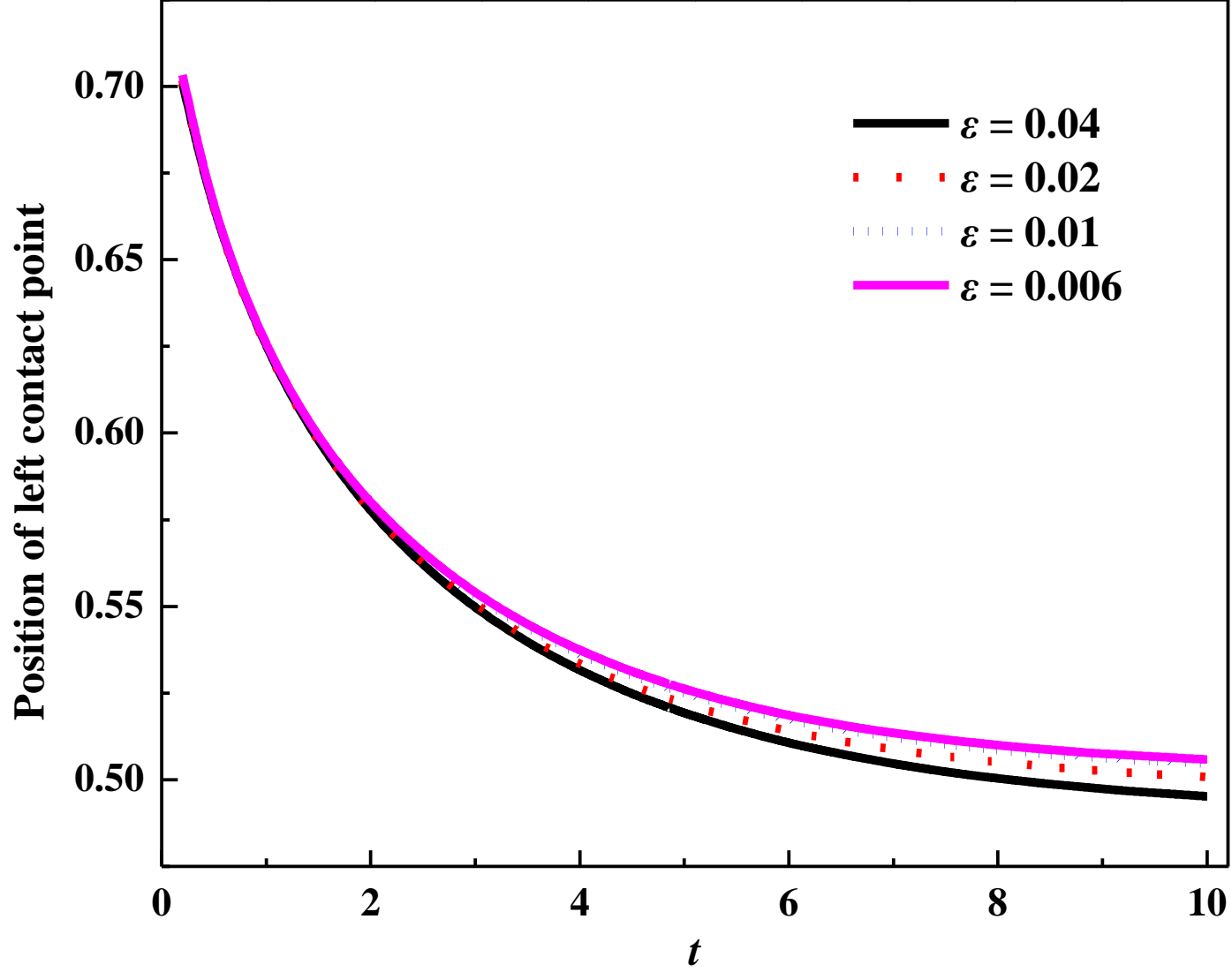

**Figure 7.** Evolutions of left contact points during the whole process.

### 5.2. Numerical validation

In this section, we use a classical benchmark to validate the proposed fully discrete scheme. The computational domain is $\Omega = [0, 1] \times [0, 2]$ with the GNBC imposed on the bottom wall ($u_w = 0$). Periodic boundary conditions are applied on the left and right sides of $\Omega$. Initially, a semicircular droplet with the radius $R_0 = 0.5$ and the contact angle $\theta_0 = 90°$ is placed on the bottom wall, as shown in Figure 8(a). The gravitational effect is neglected in this case. The droplet driven by the unbalanced Young stress will spread or recoil to the equilibrium shape with the prescribed equilibrium contact angle $\theta_e$. In equilibrium, the spreading length $L$ and droplet height $H$ in Figure 8(b) can be analytically obtained by the law of mass conservation [74]. We conduct several simulations in a wide range of surface wettability for both hydrophilic and hydrophobic cases ($\theta_e$ varies from 45° to 135°). We use a grid size of 320 × 160 and time step-size $\delta t = 5\times10^{-4}$ in all simulations. Other parameters are same as the case in Section 5.1. Numerical spreading length $L$ and droplet height $H$ agree well with analytical results obtained by equation (5.2) for all values of $\theta_e$ in Figure 9.

$$L = 2R_0\sqrt{\frac{\pi}{2\left(\theta_e - \sin\theta_e\cos\theta_e\right)}}\sin\theta_e, \quad H = R_0\sqrt{\frac{\pi}{2\left(\theta_e - \sin\theta_e\cos\theta_e\right)}}\left(1-\cos\theta_e\right). \tag{5.2}$$

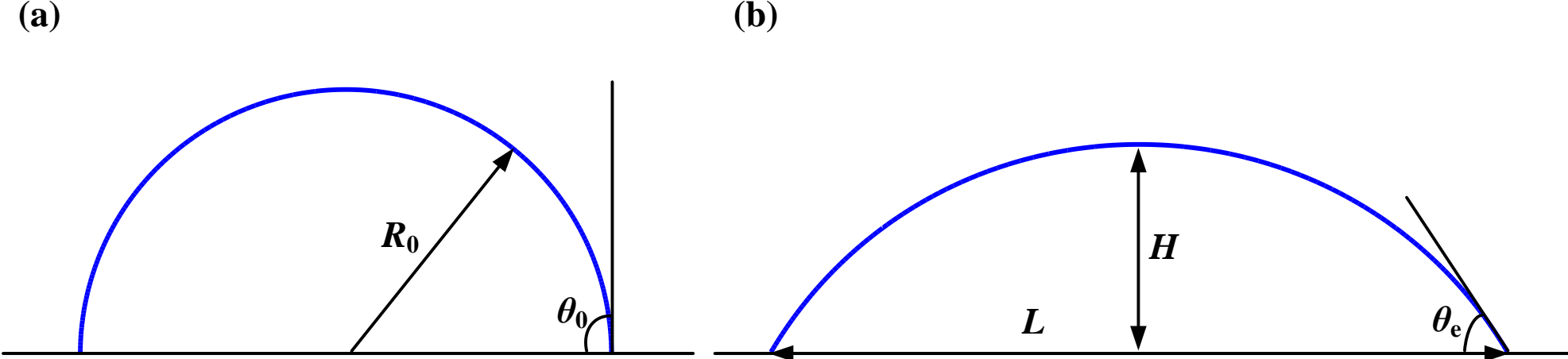

**Figure 8.** Illustration of (a) initial shape of the droplet with the radius of $R_0$ released on the bottom surface and (b) equilibrium shape of the droplet.

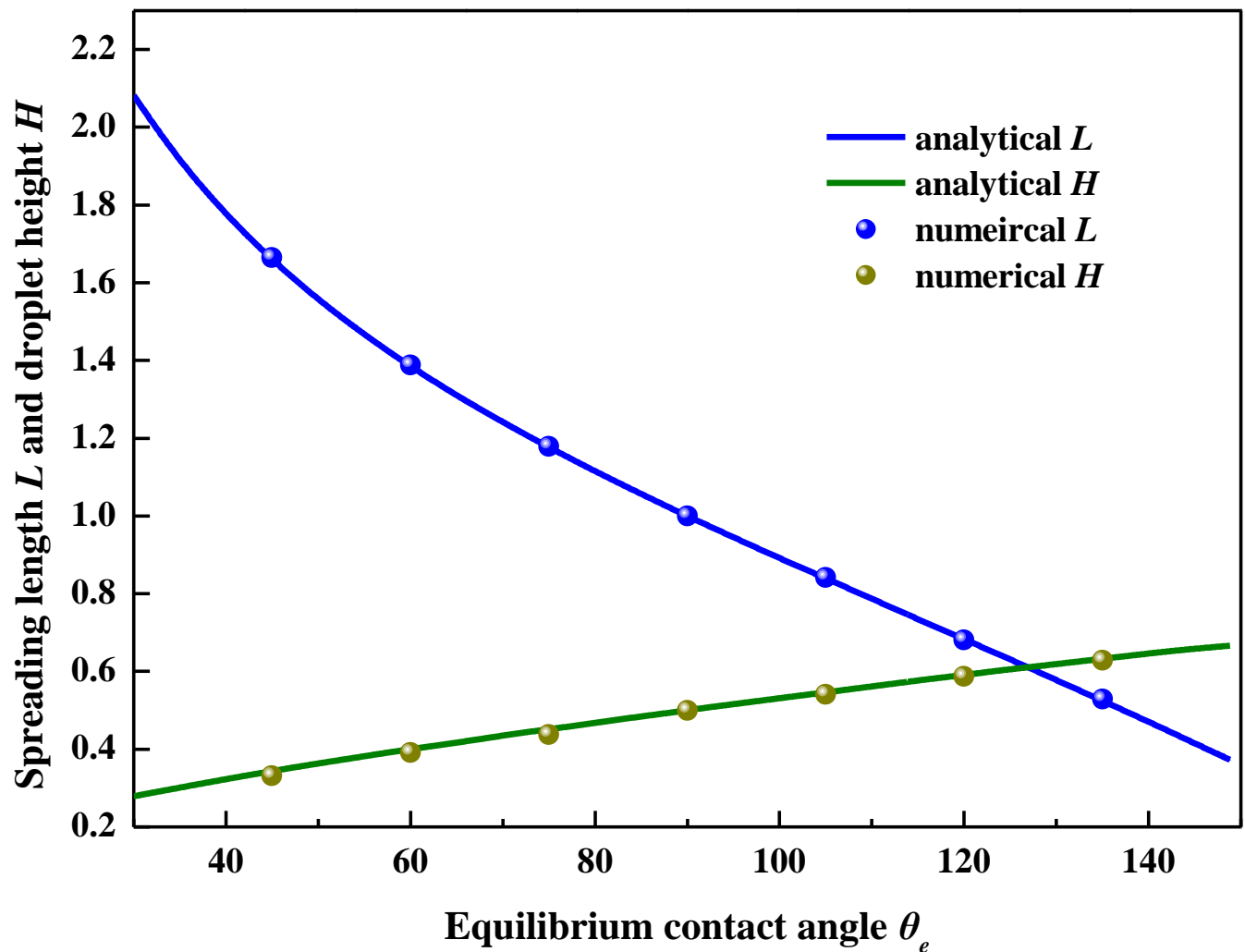

**Figure 9.** The comparison of analytical and numerical values of spreading length $L$ and droplet height $H$ at different equilibrium contact angles $\theta_e$. $\theta_e$ ranges from 45 °to 135 °with an interval 15 °.

The numerical solution will gradually converge to the specific solution if we decrease the time step-size and the grid size. Here we use the droplet spreading case to check the convergence of the decoupled scheme. A series of spatial resolutions $n_x \times n_y = 200 \times 100$, $300 \times 150$, $400 \times 200$ and $500 \times 250$ are used in this study. The time step-size $\delta t = 0.1\ h$. Evolutions of right contact points during the droplet spreading process are presented in Figure 10. It can be observed that the curve converges to a unique shape as the time step-size and the grid size decreasing.

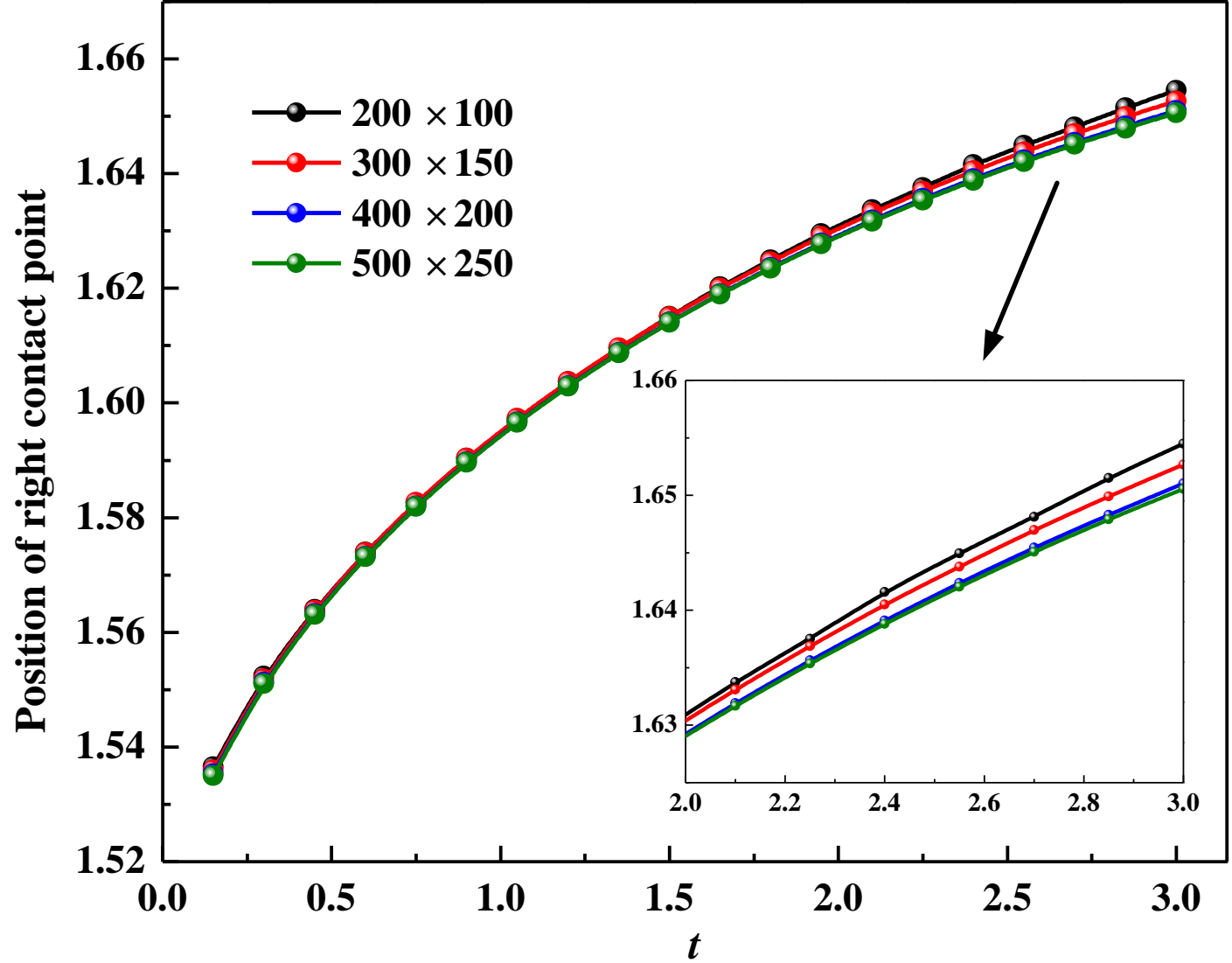

**Figure 10.** Evolutions of right contact points at different spatial resolutions.

### 5.3. Contact line dynamics on a chemically patterned surface

We use a droplet sliding case to study the contact line dynamics on a chemically patterned surface. The computational domain is $\Omega = [0, 3] \times [0, 1]$ with periodic boundary conditions applied on the left and right sides. A shear flow is imposed by moving he top wall towards the right side with a constant velocity $u_w = 1$, and the bottom surface is fixed. Initially, a semicircular droplet with the radius $R_0 = 0.5$ locates at (1, 0). The gravitational effect is neglected in this case. The bottom surface is patterned by black hydrophobic ($\theta_s = 135°$) and green hydrophilic ($\theta_s = 45°$) stripes. Two different density ratios ($\lambda_\rho$ = 1:1 and 100:1, $\lambda_\rho = \rho_1/\rho_2$) are tested in this case. The droplet and the surround fluid have the same viscosities $\eta_1 = \eta_2 = 1$. Other parameters are same as the case in section 5.1. A spatial resolution 600 × 150 and the time step-size $\delta t = 5 \times 10^{-4}$ are used in all simulations. We run each simulation until $t = 10$.

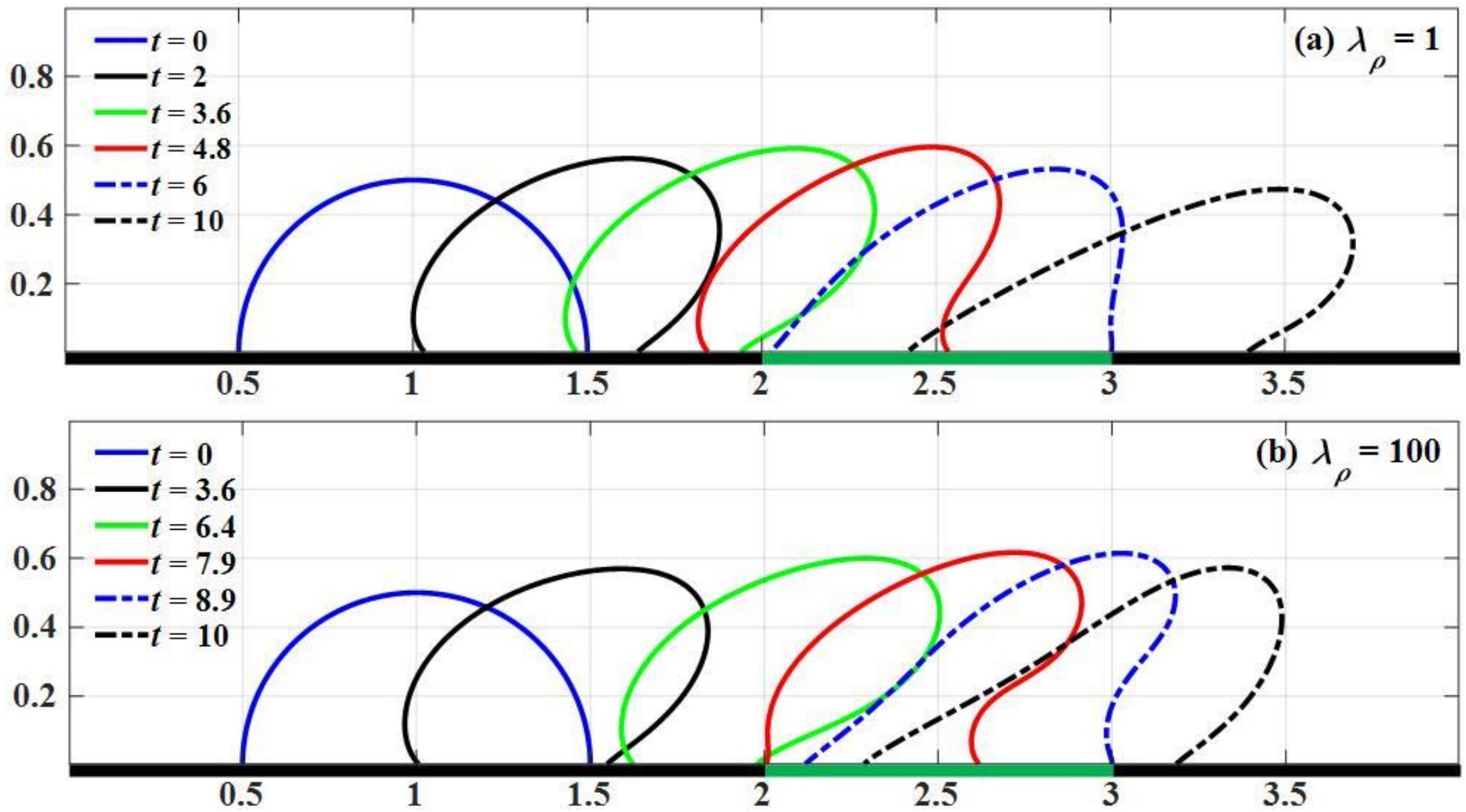

**Figure 11.** Evolutions of droplet profiles under a shear flow. Black stripe: the hydrophobic surface with the static contact angle $\theta_s = 135°$; green stripe: the hydrophilic surface with the static contact angle $\theta_s = 45°$.

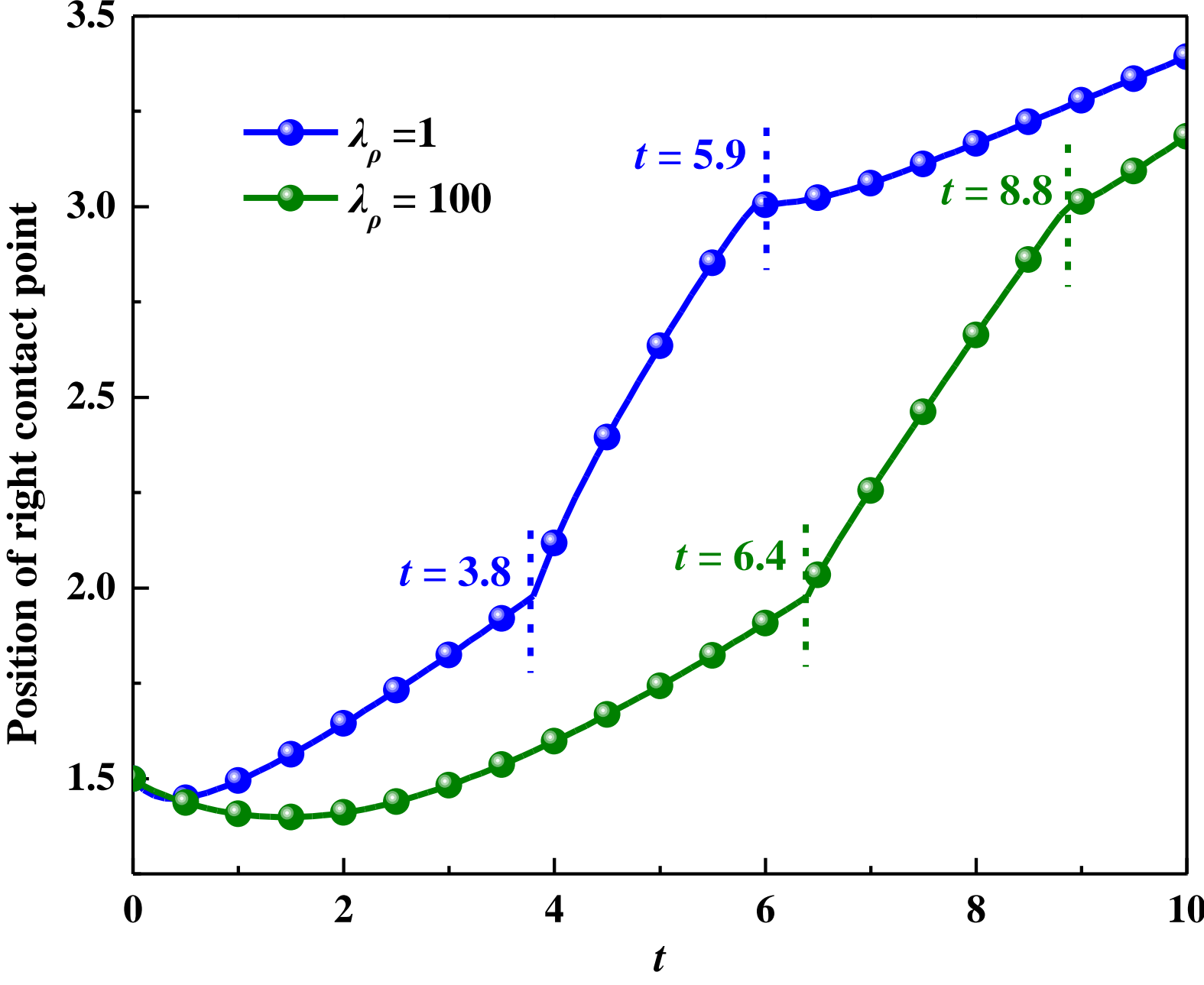

**Figure 12.** Evolutions of right contact points. Blue: $\lambda_\rho = 1$; Green $\lambda_\rho = 100$.

Figure 11 gives evolutions of droplet profiles and Figure 12 presents evolutions of right contact points during the whole process. The droplet first contracts inward on the hydrophobic stripe and then moves forward under a shear flow. For the case of $\lambda_\rho = 1$, the interface reaches the green hydrophilic stripe at $t = 3.8$, the contact line undergoes a significant acceleration followed by a deceleration (Figure 13), quickly slipping into the green stripe (Figure 14a). The capillary force dominates the adjustment of the interface profile. The opposite occurs when the interface crosses the boundary from the hydrophilic stripe to the hydrophobic stripe ($t = 5.9$), and the contact line velocity decreases almost to zero (Figure 13). As we expected, the phenomenon of the contact line sticking to the solid surface is observed (Figure 14b). We can see the similar slip-stick motion [1] of the contact line in the case of $\lambda_\rho = 100$.

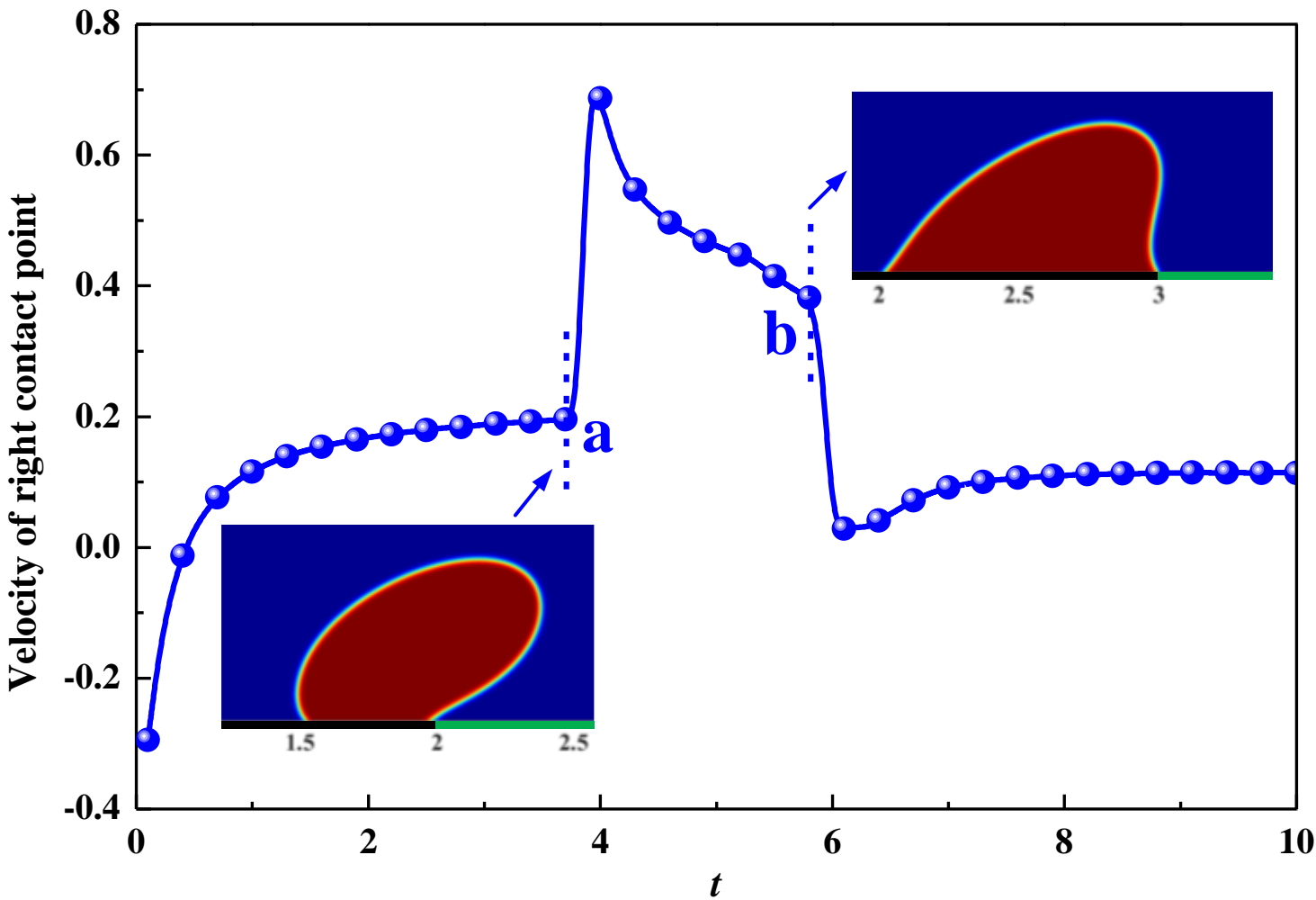

**Figure 13.** Velocity of right contact point for the case of $\lambda_\rho = 1$. At the point a, the contact line velocity increases rapidly when the interface crosses the boundary from the hydrophobic stripe to the hydrophilic stripe; The contact line velocity decreases almost to zero as the interface reaches the hydrophobic stripe (point b).

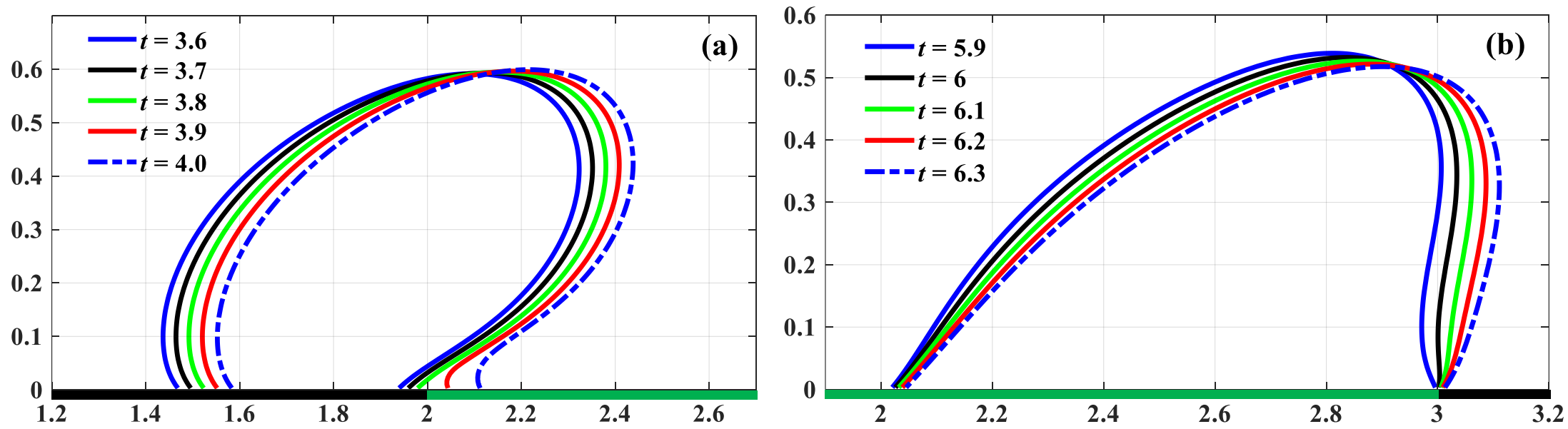

**Figure 14.** Evolutions of droplet profiles. (a) the interface moves from the black hydrophobic stripe and enters into the green hydrophilic stripe; (b) the interface crosses the stripe boundary from the green hydrophilic stripe to the black hydrophobic stripe.

### 5.4. 3D droplet spreading on a chemically patterned surface

In this section, three-dimensional simulations of droplet spreading on a fixed chemically patterned surface are conducted to demonstrate the numerical efficiency of our method. The computational domain is $\Omega = [0, 0.8]^2 \times [0, 0.4]$ with the GNBCs imposed on the bottom and top walls. A hemispherical droplet initially locates at the center of bottom wall with the radius $r = 0.25$.

On the bottom wall, we set the wetting angle $\theta_1$ for the subdomain (gray zone) $\Omega_1=\{(x, y, 0)\in \mathrm{R}^3, 0.3\le x\le 0.5, 0.3\le y\le 0.5\}$ and $\theta_2$ for anywhere else (white zone). If not explicit specified, parameters used in simulations take default values as follows:

$\rho_1 = 0.9$, $\rho_2 = 1$, $\eta_1 = 1.1$, $\eta_2 = 1$, $M_\phi = 1\times 10^{-3}$, $\varepsilon = 0.012$, $\lambda = 1.2$, $\gamma = 500$, $\beta = 5.26$, $n_x = n_y = 80$, $n_z = 40$, $\delta t = 0.1\, h_x$

We consider two different scenarios: (a) $\theta_1 = 60°$ and $\theta_2 = 120°$; (b) $\theta_1 = 120°$ and $\theta_2 = 60°$. Each simulation is run until the droplet reaches the steady state.

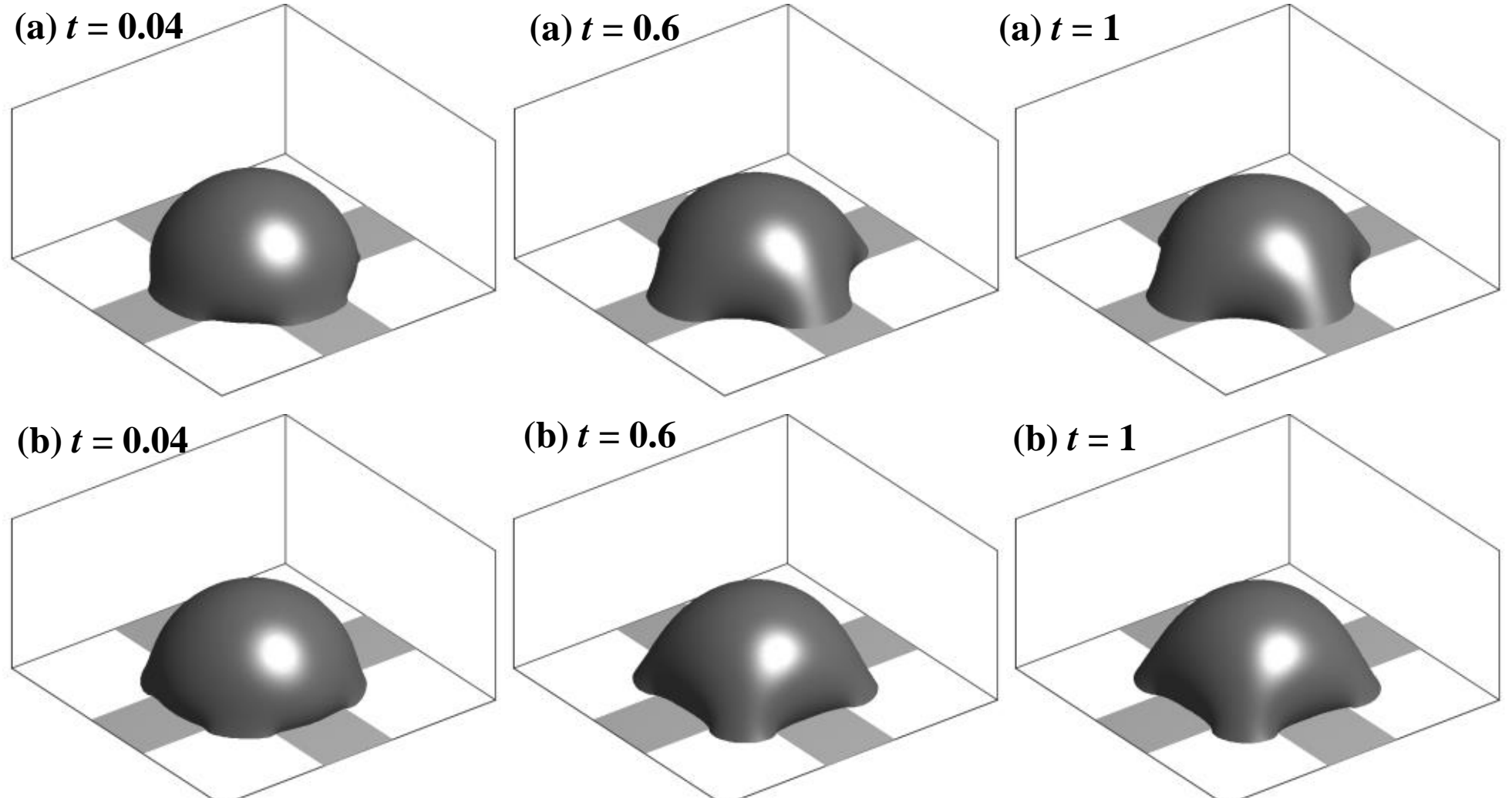

**Figure 15**. Evolutions of droplets on chemically patterned surfaces (a) $\theta_1 = 60°$ (gray zone) and $\theta_2 = 120°$ (white zone); (b) $\theta_1 = 120°$ (gray zone) and $\theta_2 = 60°$ (white zone).

Evolutions of droplets on chemically patterned surfaces are presented in Figure 15. At $t = 1$, droplets almost reach equilibrium (the total energy $E_{tot}$ reaches a minimum and no longer changes) on both surfaces. As we expected, the droplet contracts inwards on a hydrophobic zone (contact angle $\theta > 90°$) and spreads outwards on a hydrophilic zone (contact angle $\theta < 90°$) until the steady state reaches. Energy curves in Figure 16 indicate the energy stability of our scheme.

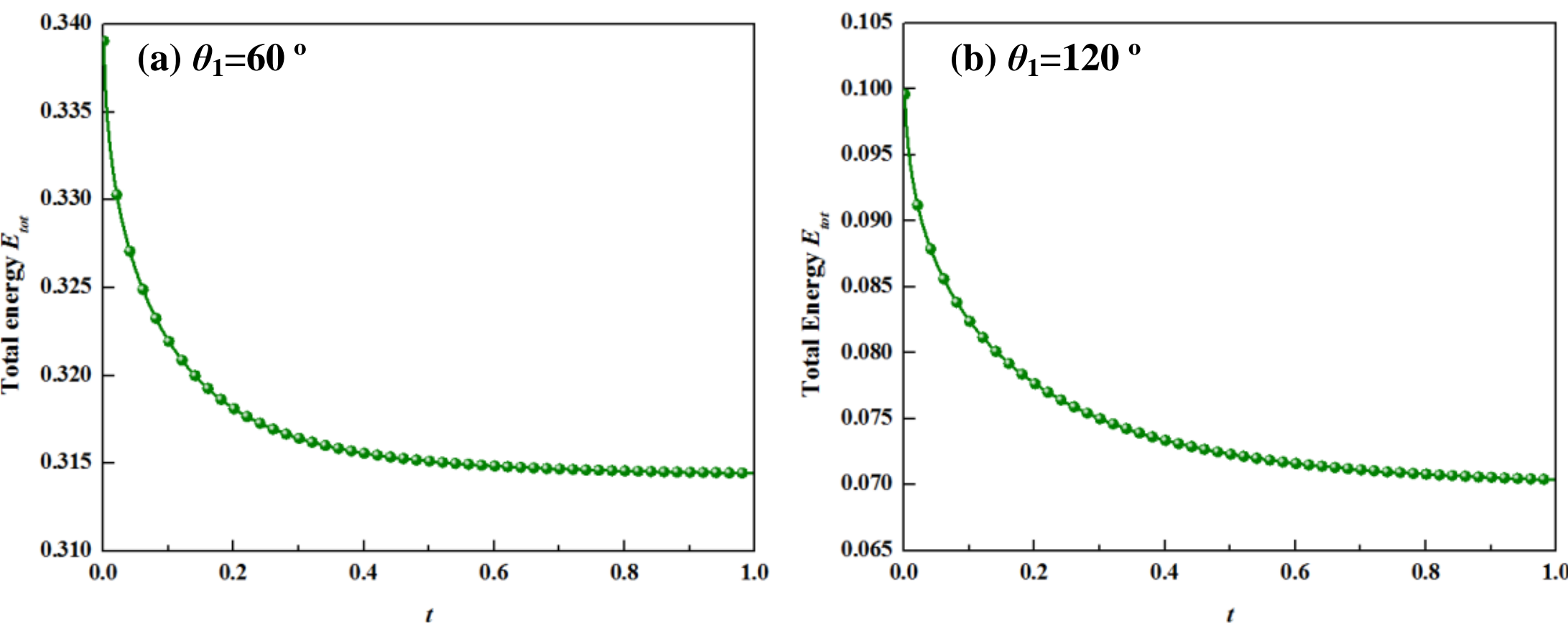

**Figure 16**. Evolutions of the total energy $E_{tot}$. (a) $\theta_1 = 60°$ (gray zone) and $\theta_2 = 120°$ (white zone); (b) $\theta_1 = 120°$ (gray zone) and $\theta_2 = 60°$ (white zone).

We further extend our simulations to a more complicated chemically patterned surface. The domain is $\Omega = [0, 0.9]^2 \times [0, 0.5]$. A hemispherical droplet is placed on the center of the patterned surface. The radius of oil droplet is 0.28. Wetting angles of the droplet on gray zones and white zones are $\theta_1 = 60°$ and $\theta_2 = 120°$, respectively. Each square box has the same size (0.1×0.1). The evolution of the droplet is presented in Figure 17.

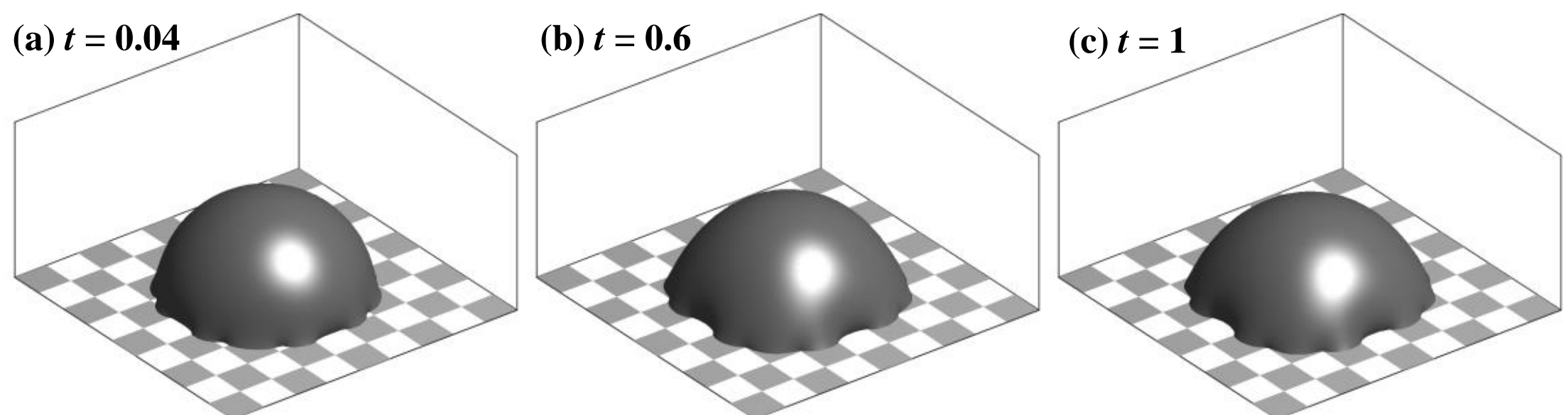

**Figure 17**. Evolution of a droplet on a chemically patterned surface with $\theta_1 = 60°$and $\theta_2 = 120°$.

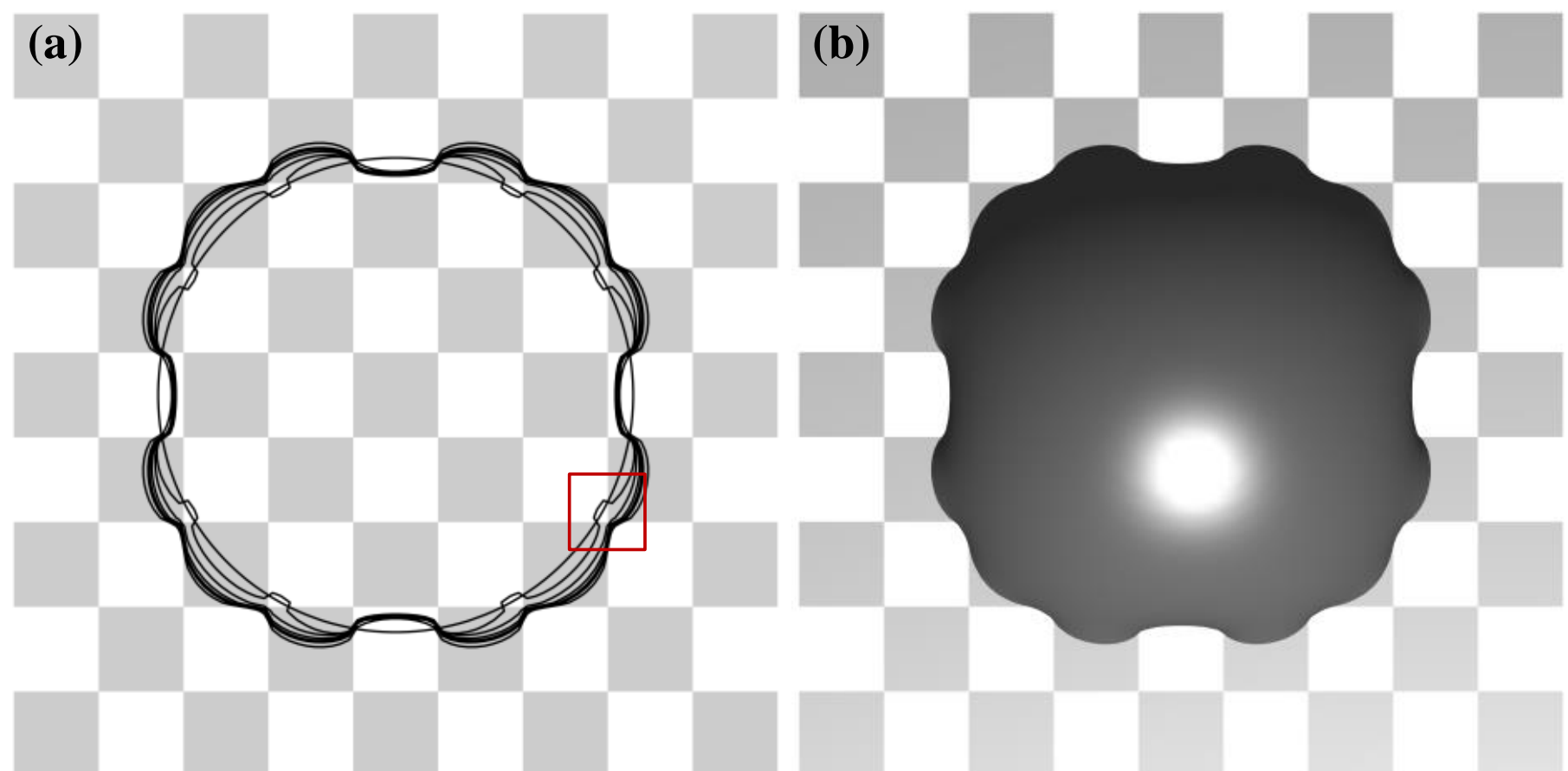

**Figure 18**. Positions of contact line from $t = 0$ to $t = 1$ and top view of the droplet at $t = 1$. (a) Positions of contact line; (b) top view of the droplet in equilibrium.

Figure 18 shows the positions of contact line from $t = 0$ to $t = 1$. In general, the droplet contracts inwards on a hydrophobic zone and spreads outwards quickly on a hydrophilic zone. We can also observe an interesting phenomenon on the corner of two white square boxes, which is marked with a red rectangle in Figure 18(a). Initially, the droplet at the corner contracts inward on the hydrophobic white box. Then the droplet on the white box moves outwards because of the spreading of the droplet on two gray boxes near the white box, and eventually the droplet reaches the corner of two white boxes when the whole system reaches a steady state.

## 6. Concluding remarks

In this paper, a fully discrete energy stable scheme is presented for the phase-field MCL model with variable densities and viscosities. The mathematical model consists of the Cahn–Hilliard equation, the Navier–Stokes equation and the GNBC for the MCL. We first construct an energy stable time-marching scheme, using the combined SAV and stabilization approach. The SAV approach is used to deal with the double well potential and a stabilization term is added to balance the explicit surface energy potential. A pressure stabilization method is used to decouple the computation of velocity and pressure. Some subtle implicit-explicit treatments are adopted to discretize convection and stress terms. We rigorously prove that the proposed time-marching scheme is unconditionally energy stable. Then a finite difference spatial discretization on staggered grids is used to spatially discretize the time-marching scheme. We further prove that the fully discrete scheme is unconditionally energy stable. Numerical results demonstrate that our scheme may be more accurate than the scheme constructed by the IEQ approach. Using our numerical scheme, we analyze the contact line dynamics through a shear flow driven droplet sliding case. 3D droplet spreading is also investigated on a chemically patterned surface. Our numerical simulation accurately predicts the expected energy evolutions and it successfully reproduces expected phenomena that the droplet contracts inwards on a hydrophobic zone and spreads outwards quickly on a hydrophilic zone.

## Acknowledgement

Guangpu Zhu and Huangxin Chen contribute equally to this work. Jun Yao and Guangpu Zhu acknowledge that this work is supported by the National Science and Technology Major Project (2016ZX05011-001), the NSF of China (51804325 and 51674280). The work of Huangxin Chen was supported by the NSF of China (Grant No. 11771363, 91630204, 51661135011), the Fundamental Research Funds for the Central Universities (Grant No. 20720180003) and Program for Prominent Young Talents in Fujian Province University. Shuyu Sun acknowledges that the research reported in this publication was supported in part by funding from King Abdullah University of Science and Technology (KAUST) through the grants BAS/1/1351-01, URF/1/2993-01, and REP/1/2879-01.

## Appendix A

In this section, we establish a rigorous proof of energy stability (**Theorem 3.2**) for the fully discrete scheme (3.37) – (3.46). The following summation-by-parts formulae are important to prove the unconditional energy stability of the fully discrete scheme.

**Lemma S1.** If $\phi \in P_h$ and $f \in U_h$, then

$$\begin{cases} h_x h_y \langle D_x \phi, f \rangle_{U_h} = -h_x h_y \langle \phi, d_x f \rangle_{P_h} + h_x h_y \langle f, \phi \rangle_{\Gamma_{ew}}, \\ \langle f, \phi \rangle_{\Gamma_{ew}} = -\left(1/h_x\right) \sum_{j=1}^{ny} A_x \phi_{1/2,\, j} f_{1/2,\, j} + \left(1/h_x\right) \sum_{j=1}^{ny} A_x \phi_{nx+1/2,\, j} f_{nx+1/2,\, j}. \end{cases} \tag{A.1}$$

**Proof.** Clearly, by the definition of $D_x$, we have

$$
\begin{aligned}
h_x h_y \langle D_x\phi, f\rangle_{U_h} &= \frac{h_x h_y}{2}\sum_{i=1}^{nx}\sum_{j=1}^{ny}\left(D_x\phi_{i+1/2,\, j} f_{i+1/2,\, j} + D_x\phi_{i-1/2,\, j} f_{i-1/2,\, j}\right)\\
&= \frac{h_y}{2}\sum_{i=1}^{nx}\sum_{j=1}^{ny}\left(\left(\phi_{i+1,\, j}-\phi_{i,\, j}\right) f_{i+1/2,\, j} + \left(\phi_{i,\, j}-\phi_{i-1,\, j}\right) f_{i-1/2,\, j}\right)\\
&= \frac{h_y}{2}\sum_{i=1}^{nx}\sum_{j=1}^{ny}\left(\phi_{i+1,\, j} f_{i+1/2,\, j} - \phi_{i,\, j} f_{i+1/2,\, j} + \phi_{i,\, j} f_{i-1/2,\, j} - \phi_{i-1,\, j} f_{i-1/2,\, j}\right)\\
&= \frac{h_y}{2}\sum_{i=1}^{nx}\sum_{j=1}^{ny}\left(\phi_{i,\, j} f_{i-1/2,\, j} - \phi_{i,\, j} f_{i+1/2,\, j} + \phi_{i,\, j} f_{i-1/2,\, j} - \phi_{i,\, j} f_{i+1/2,\, j}\right)\\
&\quad -\frac{h_y}{2}\sum_{j=1}^{ny}\phi_{0,\, j} f_{1/2,\, j} - \frac{h_y}{2}\sum_{j=1}^{ny}\phi_{1,\, j} f_{1/2,\, j} + \frac{h_y}{2}\sum_{j=1}^{ny}\phi_{nx,\, j} f_{nx+1/2,\, j} + \frac{h_y}{2}\sum_{j=1}^{ny}\phi_{nx+1,\, j} f_{nx+1/2,\, j},
\end{aligned}
$$

where $\phi_{0,j}$ and $\phi_{nx+1,j}$ are the values of ghost cells. Then by the definition of $A_x$, we can easily derive

$$
\begin{aligned}
& h_x h_y \langle D_x\phi, f\rangle_{U_h}\\
&= -h_x h_y \sum_{i=1}^{nx}\sum_{j=1}^{ny}\left(\phi_{i,\, j}\frac{f_{i+1/2,\, j}-f_{i-1/2,\, j}}{h_x}\right) - h_y\sum_{j=1}^{ny} A_x\phi_{1/2,\, j} f_{1/2,\, j} + h_y\sum_{j=1}^{ny} A_x\phi_{nx+1/2,\, j} f_{nx+1/2,\, j}\\
&= -h_x h_y \langle \phi, d_x f\rangle_{P_h} - h_y\sum_{j=1}^{ny} A_x\phi_{1/2,\, j} f_{1/2,\, j} + h_y\sum_{j=1}^{ny} A_x\phi_{nx+1/2,\, j} f_{nx+1/2,\, j}.
\end{aligned}
$$

Now we can conclude the desired result. □

Similarly, we can derive the other summation-by-parts formulae. If $\phi \in P_h$ and $f \in V_h$, then

$$
\begin{cases}
h_x h_y \langle D_y\phi, f\rangle_{V_h} = -h_x h_y \langle \phi, d_y f\rangle_{P_h} + h_x h_y \langle f, \phi\rangle_{\Gamma_{ns}},\\
\langle f, \phi\rangle_{\Gamma_{ns}} = -\left(1/h_y\right)\sum_{i=1}^{nx} A_y\phi_{i,\, 1/2} f_{i,\, 1/2} + \left(1/h_y\right)\sum_{i=1}^{nx} A_y\phi_{i,\, ny+1/2} f_{i,\, ny+1/2}.
\end{cases}
\tag{A.2}
$$

If $f \in N_h$, and $g \in U_h$, then

$$
\begin{cases}
h_x h_y \langle d_y f, g\rangle_{U_h} = -h_x h_y \langle f, D_y g\rangle_{N_h} + h_x h_y \langle f, g\rangle_{\Gamma_{ns}},\\
\langle f, g\rangle_{\Gamma_{ns}} = -\left(1/h_y\right)\sum_{i=1}^{nx+1} A_y g_{i-1/2,\, 1/2} f_{i-1/2,\, 1/2} + \left(1/h_y\right)\sum_{i=1}^{nx+1} A_y g_{i-1/2,\, ny+1/2} f_{i-1/2,\, ny+1/2}.
\end{cases}
\tag{A.3}
$$

If $f \in N_h$ and $g \in V_h$, then

$$
\begin{cases}
h_x h_y \langle d_x f, g\rangle_{V_h} = -h_x h_y \langle f, D_x g\rangle_{N_h} + h_x h_y \langle f, g\rangle_{\Gamma_{ew}},\\
\langle f, g\rangle_{\Gamma_{ew}} = -\left(1/h_x\right)\sum_{j=1}^{ny+1} A_x g_{1/2,\, j-1/2} f_{1/2,\, j-1/2} + \left(1/h_x\right)\sum_{j=1}^{ny+1} A_x g_{nx+1/2,\, j-1/2} f_{nx+1/2,\, j-1/2}.
\end{cases}
\tag{A.4}
$$

**Lemma S2.** For any variables $\left(u^n, \tilde{u}^{n+1}\right) \in U_h$ and $\left(v^n, \tilde{v}^{n+1}\right) \in V_h$, let $\left(u^n, v^n\right)$ be denoted by $\mathbf{u}^n$. If $\mathbf{u}^n \cdot \mathbf{n} = 0$ on boundaries, we have

$$
\begin{aligned}
&\langle u^n D_x\left(A_x\tilde{u}^{n+1}\right) + A_x\left(A_y v^n\right) d_y\left(A_y\tilde{u}^{n+1}\right), \tilde{u}^{n+1}\rangle_{U_h} + \langle A_x\left(A_y u^n\right) d_x\left(A_x\tilde{v}^{n+1}\right) + v^n D_y\left(A_y\tilde{v}^{n+1}\right), \tilde{v}^{n+1}\rangle_{V_h}\\
&+\frac{1}{2}\langle\left(D_x\left(A_x u^n\right) + d_y\left(A_x v^n\right)\right)\tilde{u}^{n+1}, \tilde{u}^{n+1}\rangle_{U_h} + \frac{1}{2}\langle\left(d_x\left(A_y u^n\right) + D_y\left(A_y v^n\right)\right)\tilde{v}^{n+1}, \tilde{v}^{n+1}\rangle_{V_h} = 0.
\end{aligned}
$$

**Proof.** Using the summation-by-parts formula (A.1), we have

$$
\begin{aligned}
&\langle u^n D_x\left(A_x\tilde{u}^{n+1}\right), \tilde{u}^{n+1}\rangle_{U_h} + \frac{1}{2}\langle D_x\left(A_x u^n\right)\tilde{u}^{n+1}, \tilde{u}^{n+1}\rangle_{U_h} \\
&= \frac{1}{2}\sum_{i=1}^{nx}\sum_{j=1}^{ny}\left(u^n_{i+1/2,\ j} D_x\left(A_x\tilde{u}^{n+1}_{i+1/2,\ j}\right)\tilde{u}^{n+1}_{i+1/2,\ j} + u^n_{i-1/2,\ j} D_x\left(A_x\tilde{u}^{n+1}_{i-1/2,\ j}\right)\tilde{u}^{n+1}_{i-1/2,\ j}\right) \\
&\quad + \frac{1}{4}\sum_{i=1}^{nx}\sum_{j=1}^{ny}\left(D_x\left(A_x u^{n+1}_{i+1/2,\ j}\right)\left(\tilde{u}^{n+1}_{i+1/2,\ j}\right)^2 + D_x\left(A_x u^{n}_{i-1/2,\ j}\right)\left(\tilde{u}^{n+1}_{i-1/2,\ j}\right)^2\right) \\
&= \frac{1}{8}\sum_{i=1}^{nx}\sum_{j=1}^{ny}\left(u^n_{i+1/2,\ j}\frac{\tilde{u}^{n+1}_{i+3/2,\ j} - \tilde{u}^{n+1}_{i-1/2,\ j}}{h_x}\left(\tilde{u}^{n+1}_{i+3/2,\ j} + \tilde{u}^{n+1}_{i-1/2,\ j}\right) + u^n_{i-1/2,\ j}\frac{\tilde{u}^{n+1}_{i+1/2,\ j} - \tilde{u}^{n+1}_{i-3/2,\ j}}{h_x}\left(\tilde{u}^{n+1}_{i+1/2,\ j} + \tilde{u}^{n+1}_{i-3/2,\ j}\right)\right) \\
&\quad + \frac{1}{8}\sum_{i=1}^{nx}\sum_{j=1}^{ny}\left(\frac{\left(u^n_{i+3/2,\ j} + u^n_{i+1/2,\ j}\right) - \left(u^n_{i+1/2,\ j} + u^n_{i-1/2,\ j}\right)}{h_x}\left(\tilde{u}^{n+1}_{i+1/2,\ j}\right)^2\right. \\
&\quad \left. + \frac{\left(u^n_{i+1/2,\ j} + u^n_{i-1/2,\ j}\right) - \left(u^n_{i-1/2,\ j} + u^n_{i-3/2,\ j}\right)}{h_x}\left(\tilde{u}^{n+1}_{i-1/2,\ j}\right)^2\right) \\
&= \frac{1}{8h_x}\sum_{i=1}^{nx}\sum_{j=1}^{ny}\left(u^n_{i+1/2,\ j}\left(\left(\tilde{u}^{n+1}_{i+3/2,\ j}\right)^2 - \left(\tilde{u}^{n+1}_{i-1/2,\ j}\right)^2\right) + u^n_{i-1/2,\ j}\left(\left(\tilde{u}^{n+1}_{i+1/2,\ j}\right)^2 - \left(\tilde{u}^{n+1}_{i-3/2,\ j}\right)^2\right)\right) \\
&\quad + \frac{1}{8h_x}\sum_{i=1}^{nx}\sum_{j=1}^{ny}\left(u^n_{i+1/2}\left(\left(\tilde{u}^{n+1}_{i-1/2,\ j}\right)^2 - \left(\tilde{u}^{n+1}_{i+3/2,\ j}\right)^2\right) + u^n_{i-1/2}\left(\left(\tilde{u}^{n+1}_{i-3/2,\ j}\right)^2 - \left(\tilde{u}^{n+1}_{i+1/2,\ j}\right)^2\right)\right) = 0.
\end{aligned}
$$

Similarly, we can derive

$$
\begin{cases}
\langle A_x\left(A_y v^n\right) d_y\left(A_y\tilde{u}^{n+1}\right), \tilde{u}^{n+1}\rangle_{U_h} + \frac{1}{2}\langle d_y\left(A_x v^n\right)\tilde{u}^{n+1}, \tilde{u}^{n+1}\rangle_{U_h} = 0, \\
\langle A_x\left(A_y u^n\right) d_x\left(A_x\tilde{v}^{n+1}\right), \tilde{v}^{n+1}\rangle_{V_h} + \frac{1}{2}\langle d_x\left(A_y u^n\right)\tilde{v}^{n+1}, \tilde{v}^{n+1}\rangle_{V_h} = 0, \\
\langle v^n D_y\left(A_y\tilde{v}^{n+1}\right), \tilde{v}^{n+1}\rangle_{V_h} + \frac{1}{2}\langle D_y\left(A_y v^n\right)\tilde{v}^{n+1}, \tilde{v}^{n+1}\rangle_{V_h} = 0.
\end{cases}
$$

Now the desired result can be easily obtained based on the above four equalities. □

Now we began to prove the unconditional energy stability of the fully discrete scheme. Using the Lemma S2, we can derive that

$$
\begin{aligned}
&\delta t\langle A_x\rho^n\left(u^n D_x\left(A_x u^{n+1}\right) + A_x\left(A_y v^n\right) d_y\left(A_y u^{n+1}\right)\right), u^{n+1}\rangle_{U_h} \\
&\quad + \delta t\langle A_y\rho^n\left(A_x\left(A_y u^n\right) d_x\left(A_x v^{n+1}\right) + v^n D_y\left(A_y v^{n+1}\right)\right), v^{n+1}\rangle_{V_h} \\
&\quad + \frac{1}{2}\delta t\langle A_x\rho^n\left(D_x\left(\rho^n A_x u^n\right) + d_y\left(A_x v^n\right)\right)u^{n+1}, u^{n+1}\rangle_{U_h} \\
&\quad + \frac{1}{2}\delta t\langle A_y\rho^n\left(d_x\left(A_y u^n\right) + D_y\left(A_y v^n\right)\right)v^{n+1}, v^{n+1}\rangle_{V_h} = 0,
\end{aligned}
$$

and

$$
\begin{aligned}
&\delta t\langle\left(J^n_x D_x\left(A_x u^{n+1}\right) + A_x\left(A_y J^n_y\right) d_y\left(A_y u^{n+1}\right)\right), u^{n+1}\rangle_{U_h} \\
&\quad + \delta t\langle\left(A_y\left(A_x J^n_x\right) d_x\left(A_x v^{n+1}\right) + J^n_y D_y\left(A_y v^{n+1}\right)\right), v^{n+1}\rangle_{V_h} \\
&\quad + \frac{1}{2}\delta t\langle\left(D_x\left(A_x J^n_x\right) + d_y\left(A_x J^n_y\right)\right), u^{n+1}\rangle_{U_h} + \frac{1}{2}\delta t\langle\left(d_x\left(A_y J^n_x\right) + D_y\left(A_y J^n_y\right)\right), v^{n+1}\rangle_{V_h} = 0.
\end{aligned}
$$

(1) Taking the discrete inner-product of (4.4) and (4.5) with $2u^{n+1}$ and $2v^{n+1}$ respectively,

$$
\begin{aligned}
&\langle \sigma_x^n u^{n+1},\, \sigma_x^n u^{n+1} \rangle_{U_h} - \langle \sigma_x^n u^{n},\, \sigma_x^n u^{n} \rangle_{U_h} + \langle \sigma_x^n \left(u^{n+1}-u^n\right),\, \sigma_x^n \left(u^{n+1}-u^n\right) \rangle_{U_h} \\
&\quad + \langle \sigma_y^n v^{n+1},\, \sigma_y^n v^{n+1} \rangle_{V_h} - \langle \sigma_y^n v^{n},\, \sigma_y^n v^{n} \rangle_{V_h} + \langle \sigma_y^n \left(v^{n+1}-v^n\right),\, \sigma_y^n \left(v^{n+1}-v^n\right) \rangle_{V_h} \\
&\quad + \langle \sigma_x^{n+1} u^{n+1},\, \sigma_x^{n+1} u^{n+1} \rangle_{U_h} - \langle \sigma_x^n u^{n+1},\, \sigma_x^n u^{n+1} \rangle_{U_h} + \langle \sigma_y^{n+1} v^{n+1},\, \sigma_y^{n+1} v^{n+1} \rangle_{V_h} \\
&\quad - \langle \sigma_y^n v^{n+1},\, \sigma_y^n v^{n+1} \rangle_{V_h} + 2\delta t \langle p^{n+1} - 2p^n + p^{n-1},\, d_x u^{n+1} \rangle_{P_h} - 2\delta t \langle p^{n+1}, d_x u^{n+1} \rangle_{P_h} \\
&\quad + 2\delta t \langle p^{n+1} - 2p^n + p^{n-1},\, d_y v^{n+1} \rangle_{P_h} - 2\delta t \langle p^{n+1}, d_y v^{n+1} \rangle_{P_h} \\
&\quad + 4\delta t \langle \sqrt{\eta^n}\, d_x u^{n+1},\, \sqrt{\eta^n}\, d_x u^{n+1} \rangle_{P_h} + 2\delta t \langle \sqrt{A_y\left(A_x \eta^n\right)}\, D_y u^{n+1},\, \sqrt{A_y\left(A_x \eta^n\right)}\, D_y u^{n+1} \rangle_{N_h} \\
&\quad + 4\delta t \langle \sqrt{A_y\left(A_x \eta^n\right)}\, D_x v^{n+1},\, \sqrt{A_y\left(A_x \eta^n\right)}\, D_y u^{n+1} \rangle_{N_h} \\
&\quad + 2\delta t \langle \sqrt{A_x\left(A_y \eta^n\right)}\, D_x v^{n+1}, \sqrt{A_x\left(A_y \eta^n\right)}\, D_x v^{n+1} \rangle_{N_h} + 4\delta t \langle \sqrt{\eta^n}\, d_y v^{n+1},\, \sqrt{\eta^n}\, d_y v^{n+1} \rangle_{P_h} \\
&\quad + 2\delta t \langle A_x \phi^n D_x w^{n+1},\, u^{n+1} \rangle_{U_h} + 2\delta t \langle A_y \phi^n D_y w^{n+1},\, v^{n+1} \rangle_{V_h} \\
&\quad - 2\delta t \langle A_y\left(A_x \eta^n\right) D_y u^{n+1},\, u^{n+1} \rangle_{\Gamma_{ns}} - 2\delta t \langle A_x\left(A_y \eta^n\right) D_x v^{n+1}, v^{n+1} \rangle_{\Gamma_{ew}} = 0,
\end{aligned} \tag{A.5}
$$

where $\sigma_x^n = \sqrt{A_x \rho^n}$ and $\sigma_y^n = \sqrt{A_y \rho^n}$.

(2) Taking the discrete inner-product of (4.6) with $2\delta t^2(p^{n+1}-2p^n+p^{n-1})/\chi$, we can derive

$$
\begin{aligned}
&\langle d_x\left(D_x\left(p^{n+1}-p^n\right)\right), \frac{2\delta t^2}{\chi}\left(p^{n+1}-2p^n+p^{n-1}\right) \rangle_{P_h} \\
&\quad + \langle d_y\left(D_y\left(p^{n+1}-p^n\right)\right), \frac{2\delta t^2}{\chi}\left(p^{n+1}-2p^n+p^{n-1}\right) \rangle_{P_h} \\
&= \frac{\delta t^2}{\chi} \langle D_x\left(p^{n+1}-2p^n+p^{n-1}\right), 2D_x\left(p^{n+1}-p^n\right) \rangle_{U_h} \\
&\quad + \frac{\delta t^2}{\chi} \langle D_y\left(p^{n+1}-2p^n+p^{n-1}\right), 2D_y\left(p^{n+1}-p^n\right) \rangle_{V_h} \\
&= -\frac{\delta t^2}{\chi}\Big( \langle D_x\left(p^{n+1}-p^n\right), D_x\left(p^{n+1}-p^n\right) \rangle_{U_h} - \langle D_x\left(p^{n}-p^{n-1}\right), D_x\left(p^{n}-p^{n-1}\right) \rangle_{U_h} \\
&\quad + \langle D_x\left(p^{n+1}-2p^n+p^{n-1}\right), D_x\left(p^{n+1}-2p^n+p^{n-1}\right) \rangle_{U_h} \Big) \\
&\quad - \frac{\delta t^2}{\chi}\Big( \langle D_y\left(p^{n+1}-p^n\right), D_y\left(p^{n+1}-p^n\right) \rangle_{V_h} - \langle D_y\left(p^{n}-p^{n-1}\right), D_y\left(p^{n}-p^{n-1}\right) \rangle_{V_h} \\
&\quad + \langle D_y\left(p^{n+1}-2p^n+p^{n-1}\right), D_y\left(p^{n+1}-2p^n+p^{n-1}\right) \rangle_{V_h} \Big) \\
&= 2\delta t \langle p^{n+1}-2p^n+p^{n-1}, d_x u^{n+1} \rangle_{P_h} + 2\delta t \langle p^{n+1}-2p^n+p^{n-1}, d_y v^{n+1} \rangle_{P_h}.
\end{aligned} \tag{A.6}
$$

(3) Taking the discrete inner-product of (4.6) with $-2\delta t^2 p^{n+1}/\chi$, we have

$$
\begin{aligned}
&\frac{\delta t^2}{\chi}\Big( \langle D_x p^{n+1}, D_x p^{n+1} \rangle_{U_h} - \langle D_x p^{n}, D_x p^{n} \rangle_{U_h} + \langle D_x\left(p^{n+1}-p^n\right), D_x\left(p^{n+1}-p^n\right) \rangle_{U_h} \Big) \\
&\quad + \frac{\delta t^2}{\chi}\Big( \langle D_y p^{n+1}, D_y p^{n+1} \rangle_{V_h} - \langle D_y p^{n}, D_y p^{n} \rangle_{V_h} + \langle D_y\left(p^{n+1}-p^n\right), D_y\left(p^{n+1}-p^n\right) \rangle_{V_h} \Big) \\
&= -2\delta t \langle p^{n+1}, d_x u^{n+1} \rangle_{P_h} - 2\delta t \langle p^{n+1}, d_y v^{n+1} \rangle_{P_h}.
\end{aligned} \tag{A.7}
$$

(4) Now summing up equations (A.6) and (A.7), we get

$$
\begin{aligned}
&\frac{\delta t^2}{\chi}\left(\langle D_x\left(p^n-p^{n-1}\right), D_x\left(p^n-p^{n-1}\right)\rangle_{U_h}+\langle D_y\left(p^n-p^{n-1}\right), D_y\left(p^n-p^{n-1}\right)\rangle_{V_h}\right)\\
&+\frac{\delta t^2}{\chi}\left(\langle D_x p^{n+1}, D_x p^{n+1}\rangle_{U_h}-\langle D_x p^n, D_x p^n\rangle_{U_h}+\langle D_y p^{n+1}, D_y p^{n+1}\rangle_{V_h}-\langle D_y p^n, D_y p^n\rangle_{V_h}\right)\\
=&\frac{\delta t^2}{\chi}\langle D_x\left(p^{n+1}-2p^n+p^{n-1}\right), D_x\left(p^{n+1}-2p^n+p^{n-1}\right)\rangle_{U_h}\\
&+\frac{\delta t^2}{\chi}\langle D_y\left(p^{n+1}-2p^n+p^{n-1}\right), D_y\left(p^{n+1}-2p^n+p^{n-1}\right)\rangle_{V_h}\\
&+2\delta t\langle p^{n+1}-2p^n+p^{n-1}, d_x u^{n+1}\rangle_{P_h}+2\delta t\langle p^{n+1}-2p^n+p^{n-1}, d_y v^{n+1}\rangle_{P_h}\\
&-2\delta t\langle p^{n+1}, d_x u^{n+1}\rangle_{P_h}-2\delta t\langle p^{n+1}, d_y v^{n+1}\rangle_{P_h}.
\end{aligned}
\tag{A.8}
$$

(5) Next, we take the difference of (4.6) at the time step $t^{n+1}$ and $t^n$ to derive

$$
\begin{aligned}
&\frac{\delta t^2}{\chi}\langle D_x\left(p^{n+1}-2p^n+p^{n-1}\right), D_x\left(p^{n+1}-2p^n+p^{n-1}\right)\rangle_{U_h}\\
&+\frac{\delta t^2}{\chi}\langle D_y\left(p^{n+1}-2p^n+p^{n-1}\right), D_y\left(p^{n+1}-2p^n+p^{n-1}\right)\rangle_{V_h}\\
\le&\ \chi\langle u^{n+1}-u^n, u^{n+1}-u^n\rangle_{U_h}+\chi\langle v^{n+1}-v^n, v^{n+1}-v^n\rangle_{U_h}\\
\le&\ \frac{1}{2}\langle \sigma_x^n\left(u^{n+1}-u^n\right), \sigma_x^n\left(u^{n+1}-u^n\right)\rangle_{U_h}+\frac{1}{2}\langle \sigma_y^n\left(u^{n+1}-u^n\right), \sigma_y^n\left(u^{n+1}-u^n\right)\rangle_{V_h}.
\end{aligned}
\tag{A.9}
$$

(6) Combining the equations (A.5), (A.8) and (A.9), we get

$$
\begin{aligned}
&\langle \sigma_x^{n+1}u^{n+1}, \sigma_x^{n+1}u^{n+1}\rangle_{U_h}-\langle \sigma_x^n u^n, \sigma_x^n u^n\rangle_{U_h}+\langle \sigma_x^n\left(u^{n+1}-u^n\right), \sigma_x^n\left(u^{n+1}-u^n\right)\rangle_{U_h}\\
&+\langle \sigma_y^{n+1}v^{n+1}, \sigma_y^n v^{n+1}\rangle_{V_h}-\langle \sigma_y^n v^n, \sigma_y^n v^n\rangle_{V_h}+\langle \sigma_y^n\left(v^{n+1}-v^n\right), \sigma_y^n\left(v^{n+1}-v^n\right)\rangle_{V_h}\\
&+\frac{\delta t^2}{\chi}\left(\langle D_x p^{n+1}, D_x p^{n+1}\rangle_{U_h}-\langle D_x p^n, D_x p^n\rangle_{U_h}+\langle D_y p^{n+1}, D_y p^{n+1}\rangle_{V_h}-\langle D_y p^n, D_y p^n\rangle_{V_h}\right)\\
&+\frac{\delta t^2}{\chi}\left(\langle D_x\left(p^n-p^{n-1}\right), D_x\left(p^n-p^{n-1}\right)\rangle_{U_h}+\langle D_y\left(p^n-p^{n-1}\right), D_y\left(p^n-p^{n-1}\right)\rangle_{V_h}\right)\\
&+4\delta t\langle \sqrt{\eta^n}d_x u^{n+1}, \sqrt{\eta^n}d_x u^{n+1}\rangle_{P_h}+2\delta t\langle \sqrt{A_y\left(A_x\eta^n\right)}D_y u^{n+1}, \sqrt{A_y\left(A_x\eta^n\right)}D_y u^{n+1}\rangle_{N_h}\\
&+4\delta t\langle \sqrt{A_y\left(A_x\eta^n\right)}D_x v^{n+1}, \sqrt{A_y\left(A_x\eta^n\right)}D_y u^{n+1}\rangle_{N_h}\\
&+2\delta t\langle \sqrt{A_x\left(A_y\eta^n\right)}D_x v^{n+1}, \sqrt{A_x\left(A_y\eta^n\right)}D_y v^{n+1}\rangle_{N_h}+4\delta t\langle \sqrt{\eta^n}d_y v^{n+1}, \sqrt{\eta^n}d_y v^{n+1}\rangle_{P_h}\\
\le&-2\delta t\langle A_x\phi^n D_x w^{n+1}, u^{n+1}\rangle_{U_h}-2\delta t\langle A_y\phi^n D_y w^{n+1}, v^{n+1}\rangle_{V_h}\\
&+2\delta t\langle A_y\left(A_x\eta^n\right)D_y u^{n+1}, u^{n+1}\rangle_{\Gamma_{ns}}+2\delta t\langle A_x\left(A_y\eta^n\right)D_x v^{n+1}, v^{n+1}\rangle_{\Gamma_{ew}}.
\end{aligned}
\tag{A.10}
$$

(7) For the boundary term in (A.10), using (4.8), we can derive

$$
\begin{aligned}
&2\delta t\langle A_y\left(A_x\eta^n\right)D_y u^{n+1}, u^{n+1}\rangle_{\Gamma_{ns}}+2\delta t\langle A_x\left(A_y\eta^n\right)D_x v^{n+1}, v^{n+1}\rangle_{\Gamma_{ew}}\\
=&\ 2\delta t\langle \sigma_{ns}\left(-\beta A_y u^{n+1}+\lambda A_x\tilde{L}^{n+1}D_x\left(A_y\phi^n\right)\right), u^{n+1}\rangle_{\Gamma_{ns}}\\
&+2\delta t\langle \sigma_{ew}\left(-\beta A_x v^{n+1}+\lambda A_y\tilde{L}^{n+1}D_y\left(A_x\phi^n\right)\right), v^{n+1}\rangle_{\Gamma_{ew}}\\
=&-2\delta t\langle \sigma_{ns}\beta A_y u^{n+1}, u^{n+1}\rangle_{\Gamma_{ns}}-2\delta t\langle \sigma_{ew}\beta A_x v^{n+1}, v^{n+1}\rangle_{\Gamma_{ew}}\\
&+2\lambda\delta t\langle \sigma_{ns}A_x\tilde{L}^{n+1}D_x\left(A_y\phi^n\right), u^{n+1} >_{\Gamma_{ns}} +2\lambda\delta t < \sigma_{ew}A_y\tilde{L}^{n+1}D_y\left(A_x\phi^n\right), v^{n+1}\rangle_{\Gamma_{ew}}.
\end{aligned}
\tag{A.11}
$$

(8) Taking the discrete inner-product of (4.1) with $2w^{n+1}$, we have

$$2\langle \phi^{n+1}-\phi^{n}, w^{n+1}\rangle_{P_h} - 2\delta t\langle A_x\phi^{n}u^{n+1}, D_x w^{n+1}\rangle_{U_h} - 2\delta t\langle A_y\phi^{n}v^{n+1}, D_y w^{n+1}\rangle_{V_h}$$
$$+ 2\delta t M_{\phi}\langle D_x w^{n+1}, D_x w^{n+1}\rangle_{U_h} + 2\delta t M_{\phi}\langle D_y w^{n+1}, D_y w^{n+1}\rangle_{V_h} = 0. \quad (A.12)$$

(9) Taking the discrete inner-product of (4.2) with $-2(\phi^{n+1}-\phi^{n})$, we have

$$-2\langle \phi^{n+1}-\phi^{n}, w^{n+1}\rangle_{P_h} + \lambda\varepsilon\Big(\langle D_x\phi^{n+1}, D_x\phi^{n+1}\rangle_{U_h} - \langle D_x\phi^{n}, D_x\phi^{n}\rangle_{U_h}$$
$$+\langle D_x\left(\phi^{n+1}-\phi^{n}\right), D_x\left(\phi^{n+1}-\phi^{n}\right)\rangle_{U_h}\Big) + \lambda\varepsilon\Big(\langle D_y\phi^{n+1}, D_y\phi^{n+1}\rangle_{V_h} - \langle D_y\phi^{n}, D_y\phi^{n}\rangle_{V_h}$$
$$+\langle D_y\left(\phi^{n+1}-\phi^{n}\right), D_y\left(\phi^{n+1}-\phi^{n}\right)\rangle_{V_h}\Big) - 2\lambda\varepsilon\langle D_y\phi^{n+1}, \phi^{n+1}-\phi^{n}\rangle_{\Gamma_{NS}}$$
$$-2\lambda\varepsilon\langle D_x\phi^{n+1}, \phi^{n+1}-\phi^{n}\rangle_{\Gamma_{EW}} + 2\lambda\langle U^{n+1}b^{n}, \phi^{n+1}-\phi^{n}\rangle_{P_h} = 0. \quad (A.13)$$

(10) Taking the discrete inner-product of (4.3) with $4\lambda U^{n+1}$, we obtain

$$\frac{2\lambda}{h_x h_y}\Big[\left(U^{n+1}\right)^2 - \left(U^{n}\right)^2 + \left(U^{n+1}-U^{n}\right)^2\Big] = 2\lambda\langle U^{n+1}b^{n}, \phi^{n+1}-\phi^{n}\rangle_{P_h}. \quad (A.14)$$

(10) For the boundary term in (A.13), applying (4.7) and (4.9) we have,

$$2\lambda\varepsilon\langle D_y\phi^{n+1},\phi^{n+1}-\phi^{n}\rangle_{\Gamma_{ns}} + 2\lambda\varepsilon\langle D_x\phi^{n+1}, \phi^{n+1}-\phi^{n}\rangle_{\Gamma_{ew}}$$
$$= 2\lambda\langle\sigma_{ns}\left(\tilde{L}^{n+1} - M'\left(\phi^{n}\right) - S\left(\phi^{n+1}-\phi^{n}\right)\right), \phi^{n+1}-\phi^{n}\rangle_{\Gamma_{ns}}$$
$$+ 2\lambda\langle\sigma_{ew}\left(\tilde{L}^{n+1} - M'\left(\phi^{n}\right) - S\left(\phi^{n+1}-\phi^{n}\right)\right), \phi^{n+1}-\phi^{n}\rangle_{\Gamma_{ew}}$$
$$= 2\lambda\delta t\langle\sigma_{ns}\tilde{L}^{n+1}, -\gamma\tilde{L}^{n+1}\rangle_{\Gamma_{ns}} + 2\lambda\delta t\langle\sigma_{ew}\tilde{L}^{n+1}, -\gamma\tilde{L}^{n+1}\rangle_{\Gamma_{ew}}$$
$$-2\lambda\delta t\langle\sigma_{ns}\tilde{L}^{n+1}, A_x\left(A_y u^{n+1}\right)d_x\left(A_x\left(A_y\phi^{n}\right)\right)\rangle_{\Gamma_{ns}}$$
$$-2\lambda\delta t\langle\sigma_{ew}\tilde{L}^{n+1}, A_y\left(A_x v^{n+1}\right)d_y\left(A_y\left(A_x\phi^{n}\right)\right)\rangle_{\Gamma_{ew}}$$
$$-2\lambda\langle\sigma_{ns}\left(M'\left(\phi^{n}\right) + S\left(\phi^{n+1}-\phi^{n}\right)\right), \phi^{n+1}-\phi^{n}\rangle_{\Gamma_{ns}}$$
$$-2\lambda\langle\sigma_{ew}\left(M'\left(\phi^{n}\right) + S\left(\phi^{n+1}-\phi^{n}\right)\right), \phi^{n+1}-\phi^{n}\rangle_{\Gamma_{ew}}. \quad (A.15)$$

By Taylor expansion $M\left(\phi\right)$, we know there exist $\zeta$ such that

$$M\left(\phi^{n+1}\right) = M\left(\phi^{n}\right) + M'\left(\phi^{n}\right)\left(\phi^{n+1}-\phi^{n}\right) + \frac{M''\left(\zeta^{n}\right)}{2}\left(\phi^{n+1}-\phi^{n}\right)^2.$$

(11) Combing equations (A.12) - (A.15) and applying the Taylor expansion, we have

$$
\begin{aligned}
&\lambda\varepsilon\left(\langle D_x\phi^{n+1}, D_x\phi^{n+1}\rangle_{U_h} - \langle D_x\phi^{n}, D_x\phi^{n}\rangle_{U_h} + \langle D_x\left(\phi^{n+1}-\phi^{n}\right), D_x\left(\phi^{n+1}-\phi^{n}\right)\rangle_{U_h}\right) \\
&+\lambda\varepsilon\left(\langle D_y\phi^{n+1}, D_y\phi^{n+1}\rangle_{U_h} - \langle D_y\phi^{n}, D_y\phi^{n}\rangle_{U_h} + \langle D_y\left(\phi^{n+1}-\phi^{n}\right), D_y\left(\phi^{n+1}-\phi^{n}\right)\rangle_{U_h}\right) \\
&+\left(\frac{2\lambda}{h_x h_y}\right)\left[\left(U^{n+1}\right)^2-\left(U^{n}\right)^2+\left(U^{n+1}-U^{n}\right)^2\right] \\
&-2\delta t\langle A_x\phi^n u^{n+1}, D_x w^{n+1}\rangle_{U_h} - 2\delta t\langle A_y\phi^n v^{n+1}, D_y w^{n+1}\rangle_{V_h} \\
&+2\delta t M_\phi\langle D_x w^{n+1}, D_x w^{n+1}\rangle_{U_h} + 2\delta t M_\phi\langle D_y w^{n+1}, D_y w^{n+1}\rangle_{V_h} \\
=&-2\lambda\delta t\langle\sigma_{ns}\tilde{L}^{n+1}, \gamma\tilde{L}^{n+1}\rangle_{\Gamma_{ns}} - 2\lambda\delta t\langle\sigma_{ew}\tilde{L}^{n+1}, \gamma\tilde{L}^{n+1}\rangle_{\Gamma_{ew}} \\
&-2\lambda\delta t\langle\sigma_{ns}\tilde{L}^{n+1}, A_x\left(A_y u^{n+1}\right)d_x\left(A_x\left(A_y\phi^n\right)\right)\rangle_{\Gamma_{ns}} \\
&-2\lambda\delta t\langle\sigma_{ew}\tilde{L}^{n+1}, A_y\left(A_x v^{n+1}\right)d_y\left(A_y\left(A_x\phi^n\right)\right)\rangle_{\Gamma_{ew}} \\
&-2\lambda\langle\sigma_{ns}\left(M\left(\phi^{n+1}\right)-M\left(\phi^{n}\right)\right), 1\rangle_{\Gamma_{ns}} - 2\lambda\langle\sigma_{ns}\left(S-\frac{M''\left(\zeta^n\right)}{2}\right), (\phi^{n+1}-\phi^n)^2\rangle_{\Gamma_{ns}} \\
&-2\lambda\langle\sigma_{ew}\left(M\left(\phi^{n+1}\right)-M\left(\phi^{n}\right)\right), 1\rangle_{\Gamma_{ew}} - 2\lambda\langle\sigma_{ew}\left(S-\frac{M''\left(\zeta^n\right)}{2}\right), (\phi^{n+1}-\phi^n)^2\rangle_{\Gamma_{ew}}.
\end{aligned}
\tag{A.16}
$$

(12) Finally, multiplying (A.10), (A.11) and (A.16) with $h_x h_y/2$, respectively, and summing up them together, we get,

$$
\begin{aligned}
&\frac{1}{2}\left\|\sigma^{n+1}\mathbf{u}^{n+1}\right\|_2^2 - \frac{1}{2}\left\|\sigma^{n}\mathbf{u}^{n}\right\|_2^2 + \lambda\left(\frac{\varepsilon}{2}\left\|\nabla\phi^{n+1}\right\|_2^2 - \frac{\varepsilon}{2}\left\|\nabla\phi^{n}\right\|_2^2\right) + \lambda\left[\left(U^{n+1}\right)^2-\left(U^{n}\right)^2\right] \\
&+\lambda\left(g\left(\phi^{n+1}\right)-g\left(\phi^{n}\right), 1\right)_\Gamma + \frac{\delta t^2}{2\chi}\left(\left\|\nabla p^{n+1}\right\|_2^2 - \left\|\nabla p^{n}\right\|_2^2\right) \\
\le& -\frac{\delta t}{2}\left\|\sqrt{\eta^n}D\left(\mathbf{u}^{n+1}\right)\right\|_2^2 - \delta t M_\phi\left\|\nabla w^{n+1}\right\|_2^2 - \delta t\beta\left\|\mathbf{u}_s^{n+1}\right\|_{\Gamma,2}^2 - \lambda\delta t\gamma\left\|\tilde{L}^{n+1}\right\|_{\Gamma,2}^2 \\
&-\lambda h_x h_y\langle\sigma_{ns}\left(S-\frac{M''\left(\zeta^n\right)}{2}\right), (\phi^{n+1}-\phi^n)^2\rangle_{\Gamma_{ns}} - \lambda h_x h_y\langle\sigma_{ew}\left(S-\frac{M''\left(\zeta^n\right)}{2}\right), (\phi^{n+1}-\phi^n)^2\rangle_{\Gamma_{ew}}.
\end{aligned}
$$

Now by the assumption of $S \ge L_1/2$, we can conclude the desired energy stability estimate. □